\documentclass[amsmath,twocolumn,amssymb,prb,aps,superscriptaddress,letterpaper,showpacs]{revtex4}
\usepackage{graphicx}
\usepackage[english]{babel}
\usepackage{amsmath}

\begin{document}
\title[A quantum dot photon entangler]{\centering Entanglement transfer from electrons\\to photons in quantum dots:\\An open quantum system approach}
\author{Jan C. Budich} 
\author{Bj{\"o}rn Trauzettel}
\address{Institute of Theoretical Physics and Astrophysics, University of W{\"u}rzburg, D-97074 W{\"u}rzburg, Germany}
\begin{abstract}
We investigate entanglement transfer from a system of two spin-entangled electron-hole pairs, each placed in a separate single mode cavity, to the photons emitted during their recombination process. Dipole selection rules and a splitting between the light-hole and the heavy-hole subbands are the crucial ingredients establishing a one-to-one correspondence between electron spins and circular photon polarizations. To account for the measurement of the photons as well as dephasing effects, we choose a stochastic Schr{\"o}dinger equation and a conditional master equation approach, respectively. The influence of interactions with the environment as well as asymmetries in the coherent couplings on the photon-entanglement is analyzed for two concrete measurement schemes. The first one is designed to violate the Clauser-Horne-Shimony-Holt (CHSH) inequality, while the second one employs the visibility of interference fringes to prove the entanglement of the photons. Because of the spatial separation of the entangled electronic system over two quantum dots, a successful verification of entangled photons emitted by this system would imply the detection of nonlocal spin-entanglement of massive particles in a solid state structure.
\end{abstract}
\maketitle
\section{Introduction}
The macroscopic scalability of its architecture is among the fundamental requirements which are to be met by any sound implementation of a quantum computer \cite{NielsenChuang}. The ability of transferring entanglement from spatially stationary building blocks, e.g. condensed matter embedded storage registers, to flying qubits, e.g. photons, would be an important step towards building such a computer. In this scenario, quantum communication via photons would replace the concept of data buses approved in the framework of classical computing. The main task of our work is to extensively investigate such a transfer process including the theoretical description of different kinds of non-idealities and measurements.\\

In the past, several proposals concerning the production of entangled photons employed the indistinguishability of two decay paths within a biexciton cascade \cite{Benson2000,GywatBiexciton,StaceMilburnBarnes,Visser2003,Troiani2006,Larque2008,Troiani2008,Pfanner2008}.
During the last years, great experimental progress has been made in this field \cite{Edamatsu2004,Ulrich2005,Stevenson2006,YoungShields2006,Greilich2006,AkopianExperimental,Avron2008,Shields2009}. In these systems, entanglement is generated during a coherent twofold decay process, whereas we intend to transfer entanglement in a controlled way from electron spins to photon polarizations. The functionality of a similar device performing the latter task has been proposed and studied in Ref.~\onlinecite{CerlettiGywat2005} using a master equation approach. However, a shortcoming of this proposal has been the necessity of postprocessing steps to disentangle the electronic system from the photons. Our system avoids this additional effort by using an entangled hole pair and an entangled electron pair instead of assuming only the electrons to be entangled (for similar proposals see Refs.~\onlinecite{TitovTrauzettel,Emary}). In this work, we go beyond existing literature and investigate the influence of dephasing in the electronic system on the entanglement of the emitted photons with the help of a quantum trajectory picture \cite{WisemanPHD}. This formally involved treatment in the framework of open quantum systems \cite{Breuer} is justified by showing unambiguously that some interesting features of the system's behaviour as to the production of entangled photons cannot be understood properly without an unravelling of the quantum master equation in terms of quantum trajectories.\\

\begin{figure}[htp]
\centering
\includegraphics[width=0.75\columnwidth]{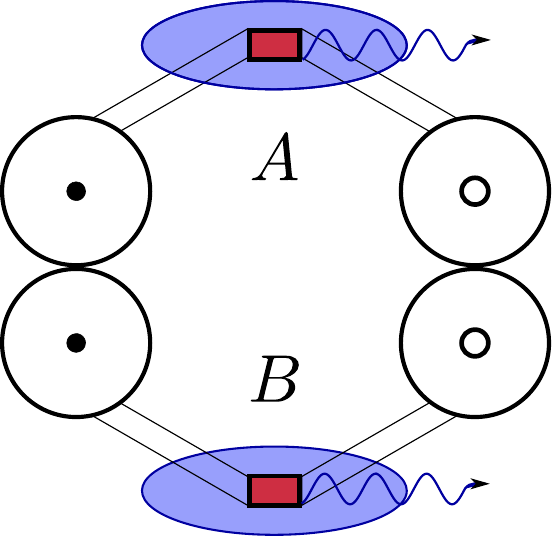}
\caption{Schematic of the photon entangler. The neighbouring circles denote lateral single dots gained by dividing double dots. The optical dots (pink boxes) contained in the nanowires (white stripes) are distinguished by naming them A and B. The black dots depict electrons, the small circles HH. Photon cavities are shown in blue oval areas around the optical dots.
}
\label{fig:entangler}
\end{figure}

In the following, we present a schematic of our entangler's functionality. A double dot device consists of a lateral quantum dot \cite{KouwenhovenReview, HansonQD} provided with an electrostatically tunable constriction potential used to divide the single dot into two separate quantum dots \cite{Blaauboer}. We assume one double dot to be charged with two electrons and another one with two heavy holes (HH). Both pairs of charge carriers are then supposed to relax into their groundstate which is a spin singlet state. By turning up the constriction potential in each of the double dots we separate the initially indistinguishable particles orbitally to yield two entangled spin pairs. Thereafter, respectively one electron and one hole are transported to an optical quantum dot via nanowires \cite{ThelanderNanowires} so that the spin pairs finally furnish two entangled exciton states delocalized over two spatially separated optical dots A and B each surrounded by a single mode cavity (see Fig.~\ref{fig:entangler}).
It will be shown that by optical recombination of the two entangled excitons a polarization entangled two photon state of the type
\begin{equation}
\lvert\Psi_f\rangle = \lvert 0 \rangle_{e^-h^+}\otimes\frac{1}{\sqrt{2}}(\lvert\sigma_A^+\sigma_B^-\rangle~ + \lvert\sigma_A^-\sigma_B^+\rangle)
\label{eqn:goalstate}
\end{equation}
can be produced which is a Bell state and thus maximally entangled. The two circularly polarized photons are then supposed to be measured after they have leaked out of the cavity surrounding dot A and dot B respectively. Preliminary experimental progress along these lines has been reported in the context of spin light-emitting diodes \cite{Fiederling1999,Ohno1999}.\\

This article is organized as follows: In Section \ref{sec:model}, a consistent modelling of the quantum dot photon entangler and its dynamics in the framework of open quantum systems will be presented. Within this model the photon entangler is assumed to be coupled weakly to its environment which exhibits the basic properties of a heat bath so that the Born-Markov approximation \cite{Blum} is applicable. A quantum stochastic differential equation (QSDE) will be derived as the equation of motion of our system of interest. The solutions of this equation are the quantum trajectories representing the time evolution of the reduced system's pure state. Furthermore, we introduce a level of description which is realistic as measured by the information a conscious observer measuring the emitted photons gains about the system state. The relevant equation of motion is then a conditional master equation \cite{StaceMilburnBarnes,WisemanPHD,GardinerNoise}. In Section \ref{sec:results}, we discuss our simulation results for the quantification of entanglement which becomes manifest in the violation of the CHSH inequality \cite{CHSH1969} and the visibility of interference fringes proving the existence of quantum coherence in a two photon Hilbert space. We work out why a treatment beyond the quantum master equation approach is adequate by capturing the fingerprints of the quantum trajectory picture in the conditional density matrix the conscious observer is aware of.
Section \ref{sec:conclusions} is aimed to sum up the most important findings and to give an outlook as to possible future work built on this article.

\section{Model \& Methods}
\label{sec:model}
\subsection{Initial state preparation}
To be capable of transferring entanglement from the electron hole excitations in the optical dots to the photon pair emitted during their recombination, the optical dots must be charged with an appropriate initial state. This may well be accomplished by taking the following steps:
\begin{itemize}
	   \item Charge each lateral quantum dot with a singlet of two electrons/HH.
	   \item Divide each lateral dot coherently into two (as shown in Fig. \ref{fig:entangler}) by turning up a constriction potential. 
	   \item Get the charge carriers transported into the optical quantum dots. 
           \item End up with two entangled exciton states delocalized over the two optical dots A and B.
	  \end{itemize}
If all these steps are taken in an absolutely coherent way, the spin state of the two $e^-$-HH excitations in the optical dots will yield
\begin{align}
\lvert\Psi_i\rangle = &\frac{1}{2}(\underbrace{\lvert\uparrow\Uparrow\rangle_A\lvert\downarrow\Downarrow\rangle_B}_{\text{dark}} - \lvert\uparrow\Downarrow\rangle_A\lvert\downarrow\Uparrow\rangle_B - \lvert\downarrow\Uparrow\rangle_A\lvert\uparrow\Downarrow\rangle_B + \nonumber\\ 
&\underbrace{\lvert\downarrow\Downarrow\rangle_A\lvert\uparrow\Uparrow\rangle_B}_{\text{dark}}),
\label{eqn:initialstate}
\end{align}
where the double-arrows denote the HH-hole spins and the simple ones represent the electrons (see Fig. \ref{fig:selectionrules}).
\subsection{Entanglement transfer due to optical dipole transitions}
\begin{figure}[htp]
\centering
\includegraphics[width=0.95\columnwidth]{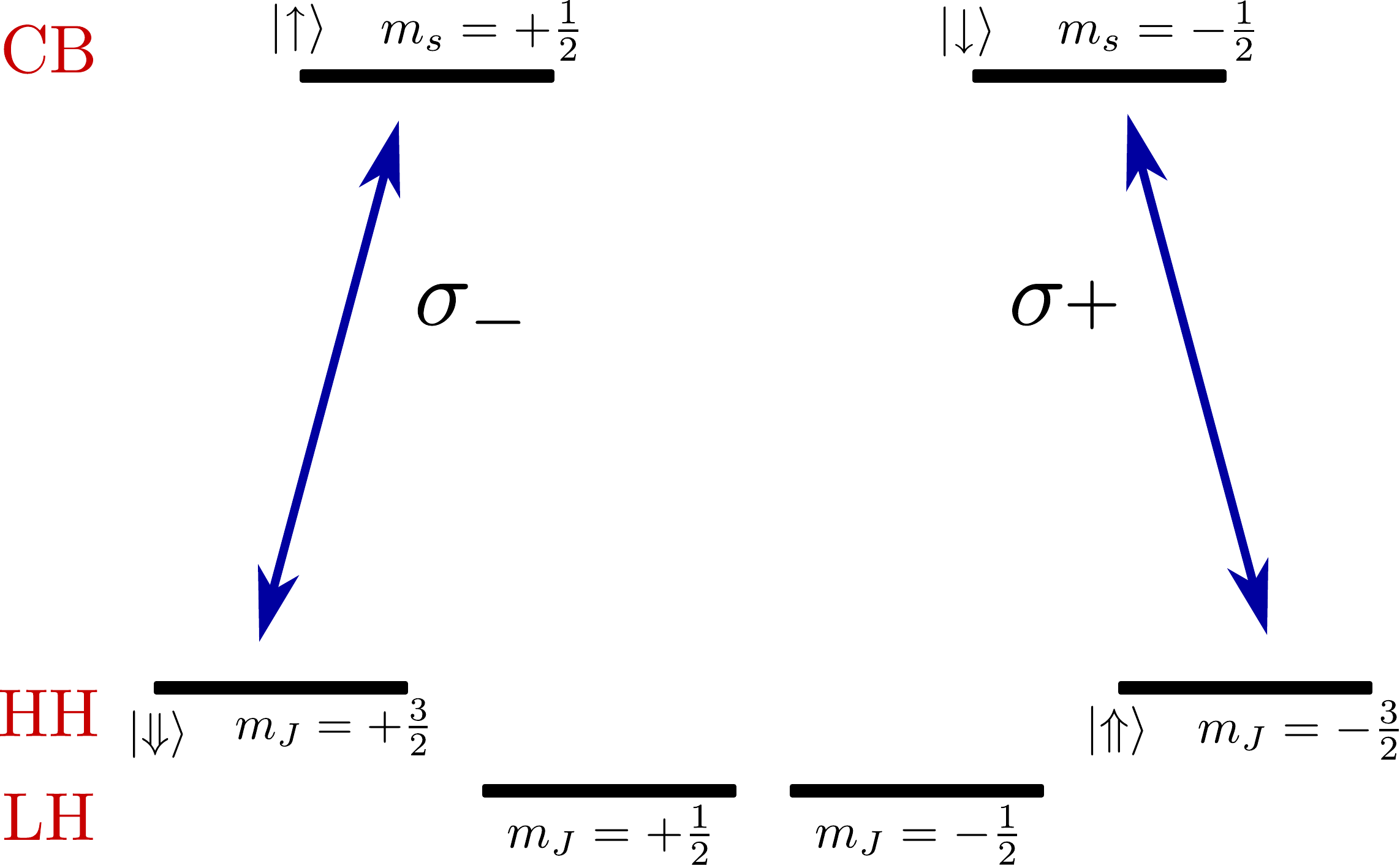}
\caption{Dipole allowed recombinations establishing a one-to-one correspondence between photon helicity and electron spin direction. $m_J = + \frac{3}{2}~$ denotes a HH-state in which a $+\frac{3}{2}$-electron is missing. The light hole (LH) levels are assumed to be far off-resonant as measured by the cavity linewidth.} 
\label{fig:selectionrules}
\end{figure}

Given the initial state of Eq. (\ref{eqn:initialstate}) our device is now ready to perform its key-task, namely the entanglement transfer from the electronic system to a photon pair. Due to the dipole selection rules $\Delta J = \pm 1,~ \Delta m_J =0,\pm 1~$ only certain recombinations of $e^-$-HH pairs are allowed (see Fig. \ref{fig:selectionrules}).
The one particle states of the $e^-$-HH excitations which are not allowed to recombine differ by 2 units of $\hbar$~in their spin z-projection. These not radiatively active states are called dark exciton states (see Eq. (\ref{eqn:initialstate})). The cavity mode for both subsystems, A and B, is assumed to be on resonance with the electron-HH recombination freqency $\omega_0$~and to propagate in z-direction. The z-direction is defined by the quantization axis of angular momentum which arises due to HH-light hole (LH) splitting. This typical splitting comes from the confinement of charge carriers in the optical dots and is amplified by mechanical strain to which the quantum dots are often exposed during the growth process. For photons emitted in z-direction, the only direction allowed by the cavity, the polarization will be circular which displays the conservation of angular momentum along the z-axis and implies $\Delta m_J = \pm 1$. In other directions the radiation of elliptically polarized light would be possible which precludes the desired one-to-one correspondence between photon helicity and electron spin projection \cite{CerlettiGywat2005}. The necessity of the HH-LH splitting becomes clear looking at Fig. \ref{fig:selectionrules} because the correspondence between circular photon polarization and electron spin would also be destroyed by transitions involving LH states. These transitions are prevented by the assumption that our cavity provides only a single radiation mode on resonance with the $e^-$-HH energy gap. However, it should be mentioned that the dark exciton states limit the fundamental efficiency of our entangler to 50\%.\\
Accounting for the selection rules and omitting the dark states our device will produce a maximally entangled two photon state (see Eq. \ref{eqn:goalstate}),
as its final state after recombination in both optical dots A and B. The dots are left in the vacuum state, i.e. without any electronic excitations and are therefore not entangled with the photons.
\subsection{System states and dynamics}
This section is intended to state more precisely the behaviour of our system. The system's Hilbert space and its coherent dynamics as well as models for environmental influences are introduced. For clarity, we start with a subsystem containing only the two optical dots before its coupling to the lateral dots provided by the nanowires is investigated.
\subsubsection{Hilbert space for the two-dot system}
We model the conduction band and the HH band with a single level for each spin projection assuming that other subbands are shifted far away, as measured by the linewidth of our cavity mode, due to spatial confinement. Furthermore, we assume that the energy levels are identical in both subsystems A and B. An overview over all states reachable from the initial state Eq. (\ref{eqn:initialstate}) is given by 
\begin{align}
  &\text{\bf States with two excitations}\nonumber\\
  &\lvert 1 \rangle = h_{\Uparrow^A}^\dag h_{\Downarrow^B}^\dag c_{\downarrow^A}^\dag c_{\uparrow^B}^\dag \lvert 0 \rangle = \lvert \Uparrow \downarrow\rangle_A\otimes\lvert \Downarrow\uparrow\rangle_B\otimes\lvert 0\rangle_{ph}\nonumber\\
  &\lvert 2 \rangle = h_{\Downarrow^A}^\dag h_{\Uparrow^B}^\dag c_{\uparrow^A}^\dag c_{\downarrow^B}^\dag \lvert 0 \rangle= \lvert \Downarrow \uparrow\rangle_A\otimes\lvert \Uparrow\downarrow\rangle_B\otimes\lvert 0\rangle_{ph}\nonumber\\
  &\lvert 3 \rangle = h_{\Downarrow^B}^\dag c_{\uparrow^B}^\dag a_{\sigma_+^A}^\dag\lvert 0 \rangle = \lvert 0\rangle_A\otimes\lvert \Downarrow\uparrow\rangle_B\otimes\lvert \sigma_A^+\rangle_{ph}\nonumber\\
  &\lvert 4 \rangle = h_{\Uparrow^B}^\dag c_{\downarrow^B}^\dag a_{\sigma_-^A}^\dag \lvert 0 \rangle=\lvert 0 \rangle_A\otimes\lvert \Uparrow\downarrow\rangle_B\otimes\lvert \sigma_A^-\rangle_{ph}\nonumber\\
  &\lvert 5 \rangle = h_{\Uparrow^A}^\dag c_{\downarrow^A}^\dag a_{\sigma_-^B}^\dag \lvert 0 \rangle=\lvert \Uparrow \downarrow\rangle_A\otimes\lvert 0\rangle_B\otimes\lvert \sigma_B^-\rangle_{ph}\nonumber\\
  &\lvert 6 \rangle = h_{\Downarrow^A}^\dag c_{\uparrow^A}^\dag a_{\sigma_+^B}^\dag \lvert 0 \rangle=\lvert \Downarrow \uparrow\rangle_A\otimes\lvert 0\rangle_B\otimes\lvert \sigma_B^+\rangle_{ph}\nonumber\\
  &\lvert 7 \rangle = a_{\sigma_+^A}^\dag a_{\sigma_-^B}^\dag\lvert 0 \rangle=\lvert 0\rangle_A\otimes\lvert 0\rangle_B\otimes\lvert \sigma_A^+\sigma_B^-\rangle_{ph}\nonumber\\
  &\lvert 8 \rangle = a_{\sigma_-^A}^\dag a_{\sigma_+^B}^\dag\lvert 0 \rangle=\lvert 0\rangle_A\otimes\lvert 0\rangle_B\otimes\lvert \sigma_A^-\sigma_B^+\rangle_{ph}\nonumber\\
  &\text{\bf States with one excitation}\nonumber\\
  &\lvert 9 \rangle = a_{\sigma_-^B}^\dag\lvert 0 \rangle=\lvert 0\rangle_A\otimes\lvert 0\rangle_B\otimes\lvert \sigma_B^-\rangle_{ph}\nonumber\\
  &\lvert 10 \rangle = h_{\Downarrow^B}^\dag c_{\uparrow^B}^\dag\lvert 0 \rangle=\lvert 0\rangle_A\otimes\lvert \Downarrow\uparrow\rangle_B\otimes\lvert 0\rangle_{ph}\nonumber\\
  &\lvert 11 \rangle =  a_{\sigma_+^A}^\dag\lvert 0 \rangle=\lvert 0\rangle_A\otimes\lvert 0\rangle_B\otimes\lvert \sigma_A^+\rangle_{ph}\nonumber\\
  &\lvert 12 \rangle = h_{\Uparrow^A}^\dag c_{\downarrow^A}^\dag\lvert 0 \rangle=\lvert \Uparrow \downarrow\rangle_A\otimes\lvert 0\rangle_B\otimes\lvert 0\rangle_{ph}\nonumber\\
  &\lvert 13 \rangle =  a_{\sigma_+^B}^\dag\lvert 0 \rangle=\lvert 0\rangle_A\otimes\lvert 0\rangle_B\otimes\lvert\sigma_B^+\rangle_{ph}\nonumber\\
  &\lvert 14 \rangle = h_{\Uparrow^B}^\dag c_{\downarrow^B}^\dag \lvert 0 \rangle=\lvert 0 \rangle_A\otimes\lvert \Uparrow\downarrow\rangle_B\otimes\lvert 0\rangle_{ph}\nonumber\\
  &\lvert 15 \rangle = a_{\sigma_-^A}^\dag\lvert 0 \rangle=\lvert 0\rangle_A\otimes\lvert 0\rangle_B\otimes\lvert \sigma_A^-\rangle_{ph}\nonumber\\
  &\lvert 16 \rangle =h_{\Downarrow^A}^\dag c_{\uparrow^A}^\dag  \lvert 0 \rangle= \lvert \Downarrow \uparrow\rangle_A\otimes\lvert 0\rangle_B\otimes\lvert 0\rangle_{ph}\nonumber\\
  &\text{\bf Vacuum state without any excitation}\nonumber\\
  &\lvert 0 \rangle = \lvert 0\rangle_A\otimes\lvert 0\rangle_B\otimes\lvert 0\rangle_{ph}. 
\label{eqn:csysstates}  
\end{align}
 The blocks are arranged by number of excitations. $a_{\sigma_i^k}^\dag,~i=\pm,~k=A,B$~creates a photon with polarization i and frequency $\omega_0$~in cavity k. $h_{\Downarrow^k},~k=A,B~$ creates a heavy hole with spin down and annihilates an electron with $m_J=+\frac{3}{2}$~in the valence band of dot k respectively. $c_{\uparrow^k}^\dag,~k=A,B$~creates an electron with $m_J = +\frac{1}{2}$~in the conduction band of dot k. Double arrows denote HH, single arrows denote electrons. Note that the dark states, the dynamics of which is irrelevant as to the production of photons, are neglected.
\subsubsection{Coherent dynamics}
The Hamiltonian of the two-dot system seen as a closed system is defined by the interaction of the electronic degrees of freedom with the radiation field which is, given a single mode cavity on resonance surrounding each optical dot, governed by the Jaynes-Cummings model \cite{JaynesCummingsOriginal}. In the interaction picture, i.e. after subtracting the noninteracting Hamiltonian which is nothing but the sum of the excitation number operators, the Hamiltonian for one optical dot reads in the rotating wave approximation
\begin{equation}
H_1^k = q_k~c_{\uparrow^k}^\dag h_{\Downarrow^k}^\dag \otimes a_{\sigma_-^k}+ v_k~c_{\downarrow^k}^\dag h_{\Uparrow^k}^\dag\otimes a_{\sigma_-^k} +h.c.;\quad k = A,B;
\label{eqn:JaynesCummingsRes}
\end{equation}
with the cavity coupling strengths $\{q_i,v_i\}~$ and the annihilation operator of the cavity mode $a_{\sigma_\pm^i}~$ of cavity i with polarization $\sigma_+~$ and $\sigma_-~$ respectively. The total interaction Hamiltonian of both dots is then given by
\begin{equation}
H = H_1^{(A)}\otimes \mathbf{1}^{(B)} + \mathbf{1}^{(A)}\otimes H_1^{(B)}.
\label{eqn:csyshamiltonian}
\end{equation}
The interaction Hamiltonian commutes with the total excitation number. The coherent unfolding is therefore not capable of connecting states of different excitation numbers (see (Eq. \ref{eqn:csysstates})). This kind of transitions can only be fulfilled by incoherent processes which will be treated next. Note furthermore that the coherent dynamics generated by H factorizes so that the total time evolution operator can be written as a tensor product, that is
\begin{equation}
U(t,t_0) = U_1^A(t,t_0)\otimes U_1^B(t,t_0) = e^{-iH_1^A (t-t_0)}\otimes e^{-iH_1^B(t-t_0)}.
\end{equation}
The following analysis will therefore be done in the subspace of one optical dot.
\subsubsection{Dissipation channels}
\begin{figure}[htp]
\centering
\includegraphics[width=0.95\columnwidth]{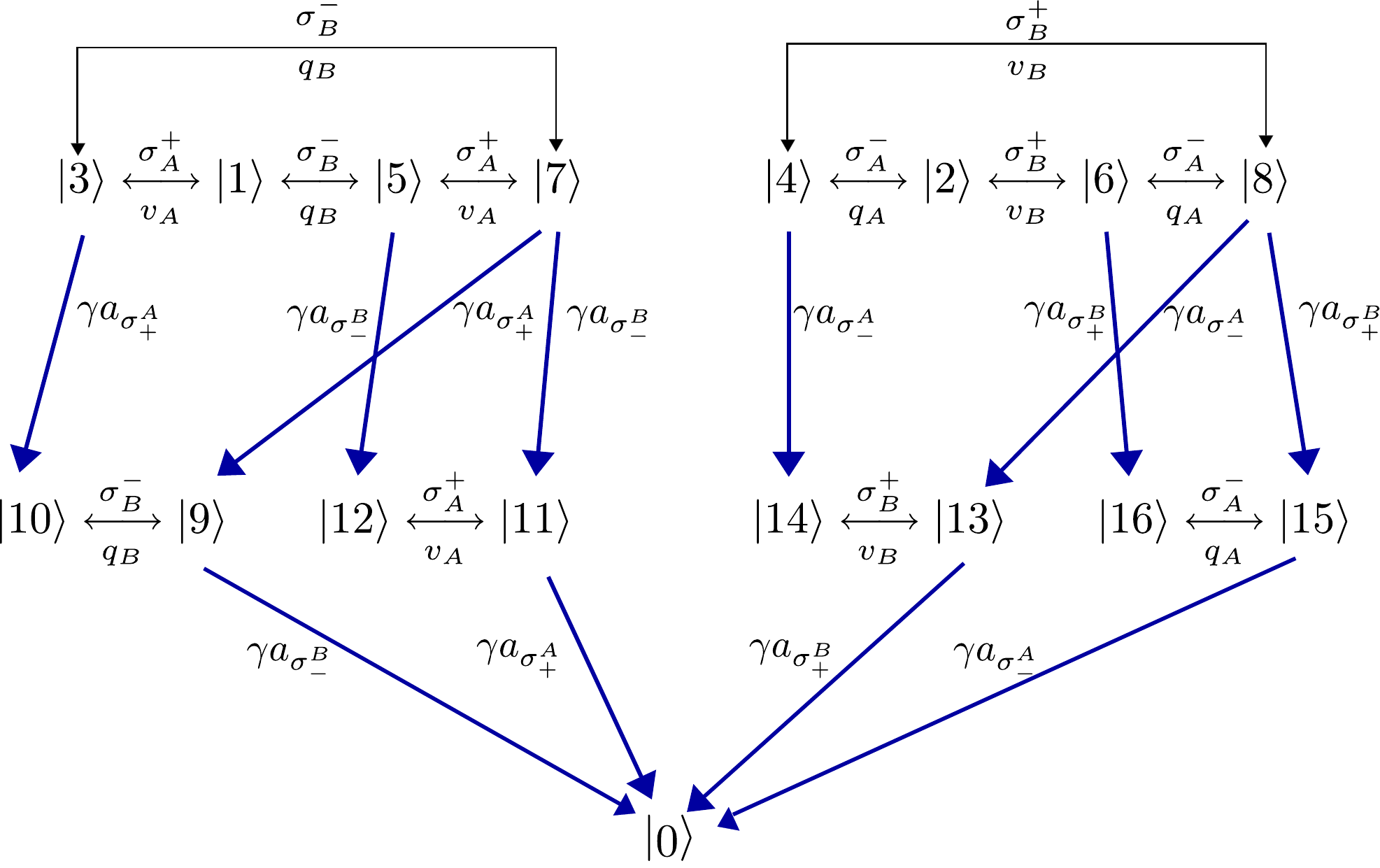}
\caption{Open quantum system dynamics of the two-dot system. The rows are arranged by excitation number. The blue, down pointing arrows represent photon measurements. Coherent couplings are drawn with black arrows. The dephasing operators are not pictured for clarity.} 
\label{fig:CSYSDynamics}
\end{figure}

Now we account for the imperfection of the cavity, which is essential for later purposes because we are only capable of measuring photons which have leaked out of the cavity. The cavity mode couples to the surrounding radiation field which is assumed to be initially in the vacuum state, i.e. in thermodynamic equilibrium at zero temperature. The corresponding dynamics is specified by the following ansatz for the interaction picture leakout Hamiltonian:
\begin{equation}
H^{L}_k(t)=e^{i\omega_0t}\left\{a_{\sigma_+^k}^\dag\otimes B^{+}_k(t) + a_{\sigma_-^k}^\dag\otimes B^{-}_k(t)\right\}+h.c.,
\end{equation}
where k = A,B and $\omega_0$~denotes the resonance frequency of the electron-HH recombination which is equal to the frequency of the cavity mode. The bath operators $B^{+}_k~$ are given by
\begin{equation}
B^{+}_k(t)=(\sum_j \xi_j e^{-i\omega_jt}b_j^{k+})
\end{equation}
with the leakout coupling strength $\xi_j$~of the cavity mode to mode j of the external field. For a definition of $B^{-}_k$~one needs to exchange + and - in the above definition of $B^{+}_k$. The $b_j^{k +/-}$~are the annihilation operators of the radiation field mode j outside the cavity k = A,B with polarization $\sigma_\pm$. We approximate the radiation field modes $\{b_j\}$~outside the cavities with a continuous spectrum with DOS $D(\omega)$. Performing the Markov approximation we can calculate the jump operators $\{A_i\}$~and their respective rates $\{\beta_i\}$~relevant for the reduced system's dynamics in the framework of continuous measurement theory \cite{WisemanPHD,Breuer}. Up to second order perturbation theory in $H_I^k := H_1^k + H^L_k(t)$, we get
\begin{align}
& A_1 = a_{\sigma_+^A},~ \beta_1 = \gamma\nonumber\\
& A_2 = a_{\sigma_-^A},~ \beta_2 = \gamma\nonumber\\
& A_3 = a_{\sigma_+^B},~ \beta_3 = \gamma\nonumber\\
& A_4 = a_{\sigma_-^B},~ \beta_4 = \gamma.
\end{align}
$\gamma = 2\pi D(\omega_0)\lvert\xi(\omega_0)\rvert^2$~ is the resonant cavity leakout rate. The dynamics of the two-dot system including the cavity leakout dissipation channels is depicted in Fig. \ref{fig:CSYSDynamics}.

Up to this point the considered dynamics is rather idealized because it does not account for any uncontrollable interaction which is able to dephase the delocalized singlet states into product states. Such an interaction will now be introduced by adding four more dissipation channels which distinguish the spin polarizations of the different electron-HH excitations. This model is equivalent to an independent dephasing of the four single particle spins since the spins occur only as part of an $e^-$-HH excitation in our dynamics. The projectors on the relevant exciton numbers, i.e. our dephasing operators read:
\begin{align}
&A_5=\hat n_{ud}^A := h_{\Uparrow^A}^\dag h_{\Uparrow^A} c_{\downarrow^A}^\dag c_{\downarrow^A}\nonumber\\
&A_6=\hat n_{du}^A := h_{\Downarrow^A}^\dag h_{\Downarrow^A} c_{\uparrow^A}^\dag c_{\uparrow^A}\nonumber\\
&A_7=\hat n_{ud}^B := h_{\Uparrow^B}^\dag h_{\Uparrow^B} c_{\downarrow^B}^\dag c_{\downarrow^B}\nonumber\\
&A_8=\hat n_{du}^B := h_{\Downarrow^B}^\dag h_{\Downarrow^B} c_{\uparrow^B}^\dag c_{\uparrow^B}
\end{align}
Note that if coherences between the two excitons occurring in the initial state get lost, neither the electrons nor the holes will be entangled any longer. This would also preclude the entanglement of the emitted photons, of course.
The dephasing jump operators $\hat n_{ud}^A, \hat n_{du}^A, \hat n_{ud}^B, \hat n_{du}^B$~are assumed to act at a rate $\beta_i=\kappa,~i=5,\ldots,8$~which we call the dephasing strength. These operators also act only on the subsystem Hilbert spaces $\mathcal H_A, \mathcal H_B$ of the two-dot system.
\subsubsection{Stochastic equations of motion}
Accounting for all dissipation channels just introduced, the quantum stochastic differential equation (QSDE) for the two-dot system is given by
\begin{equation}
d\Psi(t)=-iG(\Psi(t))dt + \sum_{k=1}^8\left ( \frac{A_k\Psi(t)}{\lVert A_k\Psi(t)\rVert}-\Psi(t)\right)dN_k(t),
\label{eqn:qsdeCSYS}
\end{equation}
with the nonlinear deterministic generator
\begin{equation}
G(\Psi) =(\underbrace{H-\frac{i}{2}\sum_{k=1}^8\beta_k A^\dag_k A_k}_{=:\hat H})\Psi + \frac{i}{2}\sum_{k=1}^8\beta_k\lVert A_k \Psi\rVert^2\Psi.
\end{equation}
The dissipation channels governed by the dephasing jump operators are of course not controllable for a conscious observer so that, if consideration is given to a realistic description, the pure state dynamics specified by the latter QSDE is no longer adequate in the presence of dephasing. A level of description, containing exactly the knowledge a conscious observer may gain by perfectly detecting the leakout photons, is provided by the following conditional master equation \cite{StaceMilburnBarnes},
\begin{equation}
\begin{split}
&d\rho_s = -i[H,\rho_s] dt + \\
&\sum_{k=1}^{4}\left\{\left(\frac{A_k\rho_s A_k^\dag}{\text{Tr}\{A_k^\dag A_k\rho_s \}}-\rho_s\right)dN_k+\gamma \text{Tr}\{A_k^\dag A_k\rho_s\}\rho_s dt \right.\\
&\left. -\frac{\gamma}{2}\{A_k^\dag A_k,\rho_s\} dt\right\}
+\sum_{k=5}^8\kappa\left\{A_k\rho_s A_k^\dag-\frac{1}{2}\{A_k^\dag A_k,\rho_s\}\right\}dt.
\end{split}
\label{eqn:condmastercsys}
\end{equation}
Setting the stochastic increments for the photon detections $\{dN_k\}~$ to zero in Eq. (\ref{eqn:condmastercsys}), we get
\begin{align}
&\frac{d\rho_c}{dt} = -i[H,\rho_c] + \sum_{k=1}^{4}\gamma \text{Tr}\{A_k^\dag A_k\rho_c\}\rho_c\nonumber\\
&+\sum_{k=5}^8\kappa A_k\rho_c A_k^\dag-\sum_{k=1}^8\frac{\beta_k}{2}\{A_k^\dag A_k,\rho_c\}.
\label{eqn:nonlinearcondmaster}
\end{align}
This evolution, conditional on the system not emitting a photon out of the cavity, takes place between the two photon detections and before the first one. It has been shown \cite{StaceMilburnBarnes} that the nonlinear term in the latter equation is only relevant for the normalization of the conditional density matrix. If we renormalize the density matrix at any given time by hand before calculating the expectation value of any observable it is thus sufficient to solve the following linear ordinary differential equation:
\begin{equation}
\frac{d\tilde \rho_c}{dt} = -i[H,\tilde \rho_c]
+\sum_{k=5}^8\kappa A_k\tilde \rho_c A_k^\dag-\sum_{k=1}^8\frac{\beta_k}{2}\{A_k^\dag A_k,\tilde \rho_c\},
\label{eqn:linearizedcondmaster}
\end{equation}
where $\tilde \rho_c$~is the unnormalized conditional density matrix.
\subsubsection{Coupling between optical and lateral dots}
\begin{figure}[htp]
\centering
\includegraphics[width=0.95\columnwidth]{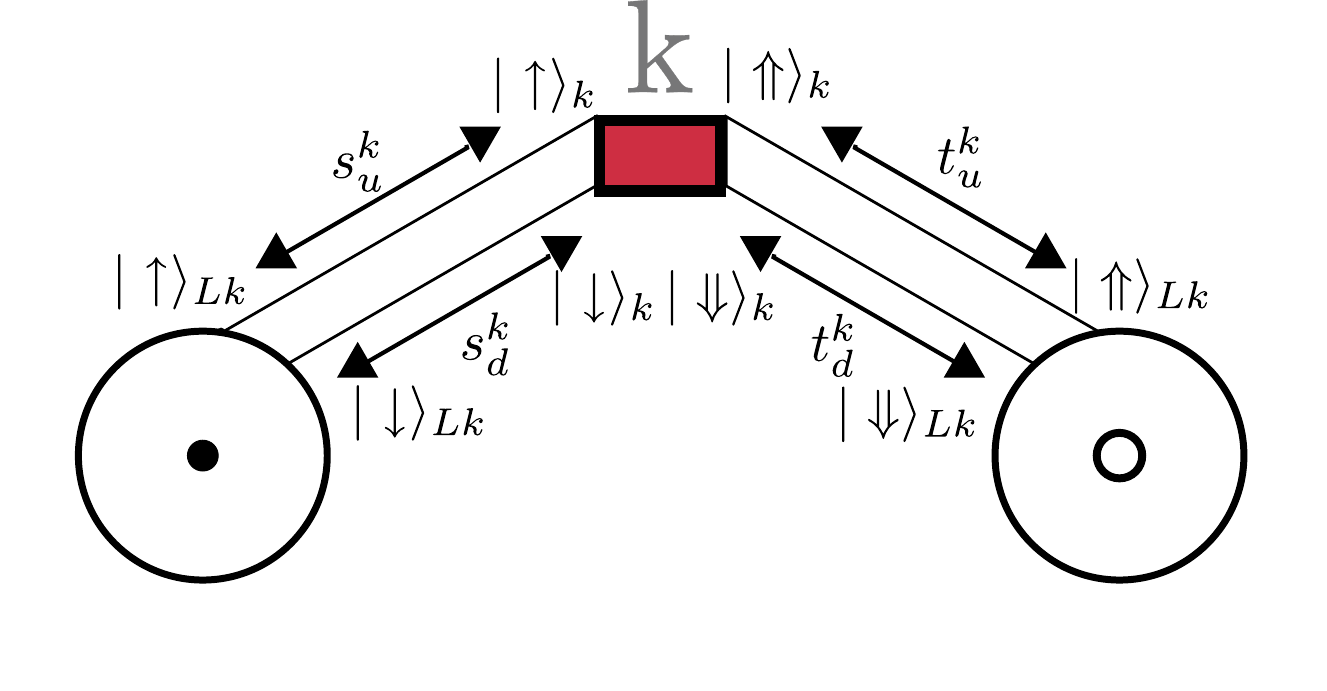}
\caption{Relevant states for the transport of electrons and HH between optical and lateral quantum dots. k = A,B. $\{s_{u/d}^k\}$~denote the tunneling rates for up/down electrons.~$\{t_{u/d}^k\}$~are the respective couplings for the HH.
} 
\label{fig:couplingstates}
\end{figure}

Concerning the transport of charge carriers from the lateral to the optical quantum dots our system can again be decoupled into two parts called A-side and B-side (see Fig.~\ref{fig:entangler}).The transport of electrons and HH through the nanowires is modelled by a simple tunneling process. Different tunneling rates for the different spin projections can be chosen to account for asymmetries of the mesoscopic wires. An A-B-dependence of the tunnel couplings models imperfections in the production of identical components. The lateral dots are treated as one level systems for each spin state. A diagram of the relevant states when consideration is given to the coupling of lateral and optical dots can be found in Fig.~\ref{fig:couplingstates}.
This description introduces eight new parameters, the four tunnel rates $\{s_{i}^k\vert i = u,d;~k = A,B\}$~for electrons and the four respective couplings $\{t_{i}^k\vert i = u,d;~k = A,B\}$~for HH.\\
The dynamics of our system is assumed to start having the particles localized in the lateral dots while the spin-state itself is not altered compared to Eq. (\ref{eqn:initialstate}). The Hamiltonian $H_S~$ for the coherent dynamics of the whole system reads
\begin{align}
H_S = H + \{&\sum_{k=A,B} s_u^k~ c_{\uparrow^k}^\dag c_{\uparrow^k}^{(L)} + s_d^k~ c_{\downarrow^k}^\dag c_{\downarrow^k}^{(L)} + t_u^k~ h_{\Uparrow^k}^\dag h_{\Uparrow^k}^{(L)} + \nonumber\\
&t_d^k~ h_{\Downarrow^k}^\dag h_{\Downarrow^k}^{(L)} + h.c.\},
\label{eqn:defhs}
\end{align}
where the index $(L)$~denotes the lateral quantum dots. In the simulations for the coupled system we neglect dephasing so that we don't need to expand our dephasing model to the enlarged system. The difference between the whole system and the two-dot system is then simply the substitution of $H_s$~for $H$~in the respective stochastic equations of motion since the cavity leakout concerns only the optical dots anyway.
\subsection{CHSH violation}
\label{sec:chshtheory}
Now, we prove and quantify the capability of our system to produce entangled photons by simulating a CHSH inequality \cite{CHSH1969} violating measurement. We focus on basic physical features which are essential to understand the advantages of the quantum trajectory picture. For simplicity we assume that the two photons are measured simultaneously, which could experimentally be realized by a post selection rule only keeping track of coincidental measurements. Furthermore we only investigate the two-dot system not describing the coupling to the lateral dots.\\
Without dephasing, our system is supposed to produce the maximally entangled two photon state $\lvert\Psi_{ph}\rangle=\frac{1}{\sqrt{2}}(\lvert\sigma_A^+\sigma_B^-\rangle~ + \lvert\sigma_A^-\sigma_B^+\rangle)~$ (see Eq. (\ref{eqn:goalstate})). An appropriate measurement scheme to verify the entanglement of the state measured in the simulation should accordingly be designed such that the CHSH inequality will be violated maximally by $\lvert\Psi_{ph}\rangle$, that is $\mathcal B(\Psi_{ph}) = \langle QS\rangle_{\Psi_{ph}}+\langle RS\rangle_{\Psi_{ph}}+\langle RT\rangle_{\Psi_{ph}}-\langle QT\rangle_{\Psi_{ph}}=2\sqrt{2}$. This condition is met by the subsequent set of observables.
\begin{align}
&Q =  Z_A,\quad &R =  -X_A\nonumber\\
&S =  \frac{-Z_B-X_B}{\sqrt{2}},\quad &T = \frac{Z_B -X_B}{\sqrt{2}}.
\label{eqn:bellsetcsys}
\end{align}
Note that Q and R are observables of subsystem A while S and T are observables of subsystem B. The combinations QS, QT, RS and RT are each measured with probability 0.25 in all simulations.\\
To calculate the expectation values of the relevant observables we have to solve the stochastic equation of motion, Eq.~(\ref{eqn:condmastercsys}) and Eq.~(\ref{eqn:qsdeCSYS}) respectively to obtain the reduced system's state at the photon measuring time $\tau$. We employ several methods to accomplish this integration:\\
A first choice providing an exact solution is the integration of Eq.~(\ref{eqn:linearizedcondmaster}) to calculate the density matrix $\rho_c(\tau)=\frac{\tilde \rho_c(\tau)}{\text{Tr}\tilde \rho_c(\tau)}$. The relevant desired expectation values are then given by
\begin{equation}
\langle XY \rangle_{\rho_c(\tau)} = \frac{\text{Tr}\{\tilde\rho_c(\tau)XY\}}{\text{Tr}\{\tilde\rho_c(\tau)\}}, \qquad X=Q,R;~Y=S,T.
\end{equation}
A second method to obtain these quantities is the numerical integration of the QSDE, Eq.~(\ref{eqn:qsdeCSYS}), by creating a stochastic ensemble of quantum trajectories (see Refs. \onlinecite{WisemanPHD,Breuer} for a detailed discussion). The expectation values of the mentioned observables are obtained determining the respective measurement results for each trajectory and then calculating the ensemble average. This approach reproduces the exact solution when the ensemble size goes to infinity.\\
A third technique is to calculate the integral representation of the propagator of the underlying piecewise deterministic process PDP with the help of a Hilbert space path integral \cite{Breuer}. By this we would calculate the statistical contribution of any plausible quantum trajectory and therefore obtain the same results as when averaging over an ensemble of infinite size. To realize the computation of the path integral we have to introduce a cutoff c which leads to the neglect of all trajectories containing more than c dephasing jumps. This approximation is reasonable since we are in a weak coupling regime where the Born-Markov approximation is applicable. Hence, we can think of $\kappa$~as small parameter which suppresses trajectories with i dephasing jumps with a prefactor of $\kappa^i$. The path integral technique then converges against the exact solution as c goes to infinity.\\
\subsection{Quantification of entanglement with a single interference setup}
Our second approach makes use of the fact that the dimension of the polarization Hilbert space $\mathcal H_{AB}^{ph}$~of the two leakout photons is reduced in virtue of optical selection rules. They forbid the emission of photon pairs with equal circular polarization which cuts down  $\mathcal H_{AB}^{ph}$~ to a two dimensional Hilbertspace $\mathcal{\tilde H}_{AB}^{ph}$~spanned by the basis $\mathcal{ \tilde M} = \{ \lvert\sigma_A^+ \sigma_B^-\rangle,\lvert \sigma_A^- \sigma_B^+\rangle\}$. The task of investigating polarization entanglement is then simply to distinguish a coherent superposition like $\alpha\lvert\sigma_A^+ \sigma_B^-\rangle + \beta \lvert\sigma_A^+ \sigma_B^-\rangle,\quad \lvert \alpha\rvert^2+\lvert \beta\rvert^2 = \text{1}~$ from an incoherent mixture $\lvert\alpha\rvert^2 \lvert\sigma_A^+ \sigma_B^-\rangle\langle \sigma_A^+ \sigma_B^-\rvert + \lvert \beta \rvert ^2\lvert\sigma_A^- \sigma_B^+\rangle\langle \sigma_A^- \sigma_B^+\rvert~$ and to determine the ratio of the amplitudes $\alpha$ and $\beta$. These properties can be characterized with the help of a single interference experiment.\\
Our interference setup is closely related to that proposed in Ref.~\onlinecite{StaceMilburnBarnes}.
\begin{figure}[h!tp]
\centering
\includegraphics[width=0.75\columnwidth]{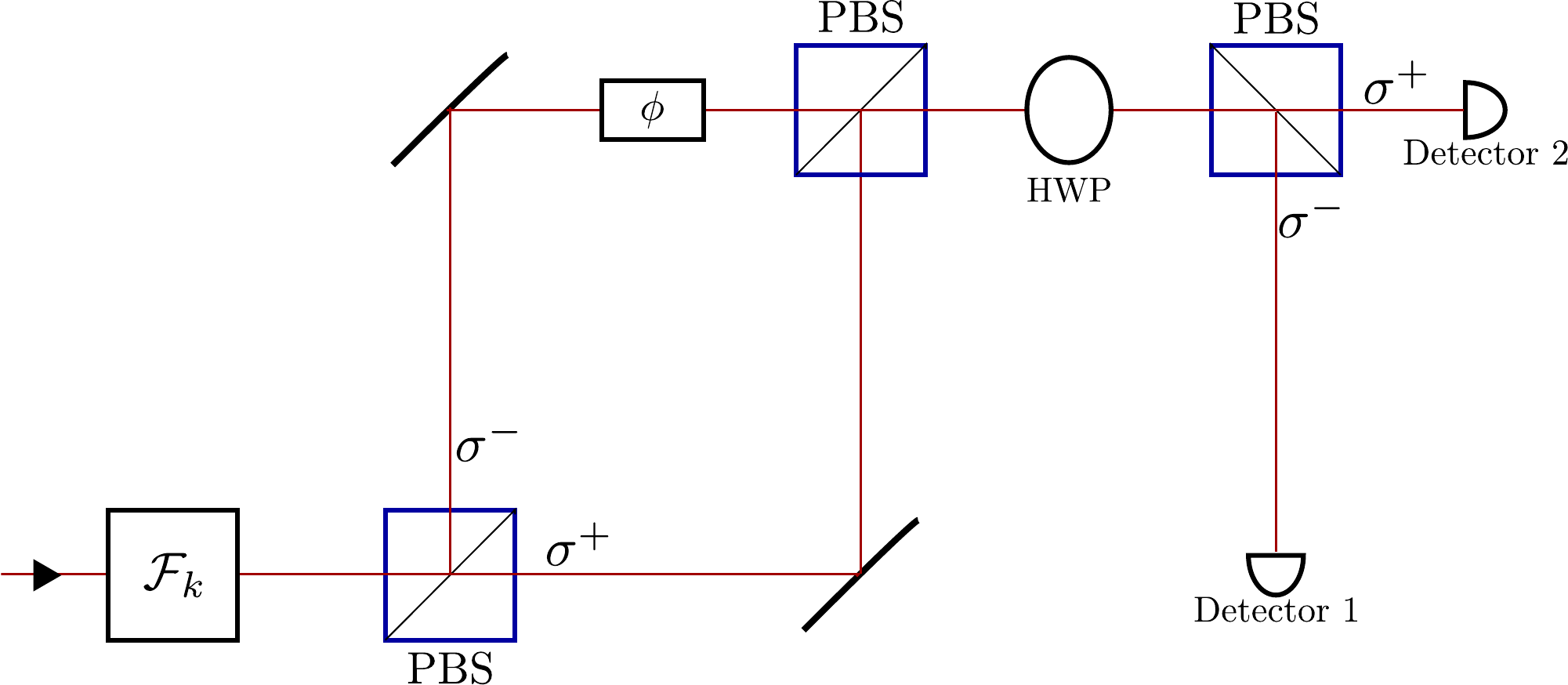}
\caption{Interference setup for the postprocessing of photons emitted by subsystem $k=A$~and~$k=B$~respectively. PBS stands for polarization beam splitter, the half wave plate is denoted with HWP. $\mathcal F_k$~performs a polarization flip on subsystem B leaving subsystem A unchanged.}
\label{fig:interferencesetup}
\end{figure} 
For later convenience we prepend a device which flips the polarization of photon B (see Fig.~\ref{fig:interferencesetup}), that is a unitary operation 
\begin{equation}
\mathcal F = \mathcal F_A\otimes\mathcal F_B = \mathbf 1_{A}^{ph}\otimes\left(\sigma_x\right)_B^{ph},
\end{equation} 
to the interferometer which without $\mathcal F$ performs the transformation
\begin{equation}
\mathcal O = \frac{1}{\sqrt{2}}\left(\begin{array}{cc} e^{i\phi}&1\\-1 & e^{-i\phi}
\end{array}\right)_A^{ph}\otimes\frac{1}{\sqrt{2}}\left(\begin{array}{cc} e^{i\phi}&1\\-1 & e^{-i\phi}
\end{array}\right)_B^{ph}.
\end{equation}
The effect of the whole setup shown in Fig.~\ref{fig:interferencesetup} on the vector of photon annihilation operators $\vec a := (a_{\sigma_+^A},a_{\sigma_-^A})^T\otimes(a_{\sigma_+^B},a_{\sigma_-^B})^T~$ reads
\begin{equation}
{\vec a}^\prime = \mathcal O \circ  \mathcal F(\vec a).
\label{eqn:interferencesubstitution}
\end{equation}
Note that the unitary operation $\mathcal O \circ  \mathcal F~$ can be factorized to a tensor product of two local operations in the respective one photon Hilbert spaces. This is of great importance, because otherwise it could alter the entanglement properties of the input state.\\
Following the analysis in Ref. \onlinecite{StaceMilburnBarnes}, we define the visibility of interference fringes $\mathcal V$~by
\begin{equation}
\mathcal V(t_B,t_A) = \frac{2\lvert Z(t_B,t_A)\rvert}{X(t_B,t_A)+Y(t_B,t_A)},
\label{eqn:visconcrete}
\end{equation}
where the quantities $X,Y,Z~$ read in our case
\begin{align}
X &= \langle: \hat n_{\sigma_+^B}(t_B) \hat n_{\sigma_-^A}(t_A):\rangle=\nonumber\\
&=\text{Tr}\left\{a_{\sigma_+^B} \mathcal T_{t_B,t_A}\left(a_{\sigma_-^A} \mathcal T_{t_A,0}\left(\rho_s(0)\right) a_{\sigma_-^A}^{\dag}\right) a_{\sigma_+^B}^{\dag}\right\}\nonumber\\
Y &= \langle: \hat n_{\sigma_-^B}(t_B) \hat n_{\sigma_+^A}(t_A):\rangle=\nonumber\\
&=\text{Tr}\left\{a_{\sigma_-^B}\mathcal T_{t_B,t_A}\left(a_{\sigma_+^A} \mathcal T_{t_A,0}\left(\rho_s(0)\right) a_{\sigma_+^A}^{\dag}\right) a_{\sigma_-^B}^{\dag}\right\}\nonumber\\
Z &= -\text{Tr}\left\{a_{\sigma_-^B}\mathcal T_{t_B,t_A}\left(a_{\sigma_+^A}\mathcal T_{t_A,0}\left(\rho_s(0)\right) a_{\sigma_-^A}^{\dag}\right) a_{\sigma_+^B}^{\dag}\right\}.
\label{eqn:xyzdef}
\end{align}
$\mathcal T_{\cdot,\cdot}$~is the conditional time evolution operator representing the flow of Eq.~(\ref{eqn:nonlinearcondmaster}), $\hat n_{\sigma_i^k},~i=\pm,~k=A,B$~are the photon number operators of the respective cavity modes and the colons denote normal ordering. The name visibility of interference fringes for $\mathcal V$~is immediately justified looking at the following equation:
\begin{equation}
\langle: \hat n_{\sigma_+^B}^\prime(t_B) \hat n_{\sigma_-^A}^\prime(t_A):\rangle = \frac{\gamma^2}{4}\left(X+Y+ Z e^{2i\phi} + Z^* e^{-2i\phi}\right).
\label{eqn:xyzinterferenceform}
\end{equation}
The one-to-one correspondence between entanglement and visibility becomes manifest in the subsequent relation between $\mathcal V~$ and the Bell parameter $\mathcal B_{max}~$ of a measurement scheme maximizing $\mathcal B~$ for a given input state \cite{StaceMilburnBarnes}:
\begin{equation}
\mathcal B_{max} = \sqrt{2}(1+\mathcal V).
\end{equation}
\section{Results \& Discussion}
\label{sec:results}
\subsection{Decay of entanglement due to dephasing}
In order to understand the basic phenomenology of entanglement decay due to dephasing we set $v_i = q_i = \text{1},~ i= A,B~$ and $\gamma = \text{0.1}~$ so that the coherent coupling of the electron-HH excitations to the cavity mode is by one order of magnitude stronger than the leakout rate and we have total symmetry between subsystems A and B. The influence of asymmetries in the model parameters is investigated with the help of the interference approach in Section \ref{sec:asymresults}. All energies are measured in units of $q_B$~throughout this work. The dimensionless timescale is then fixed by the inverse of this energy-unit.
\subsubsection{Results for the CHSH violation}
\label{sec:chshresults}
\begin{figure}[h!tp]
\centering
\includegraphics[width=0.95\columnwidth]{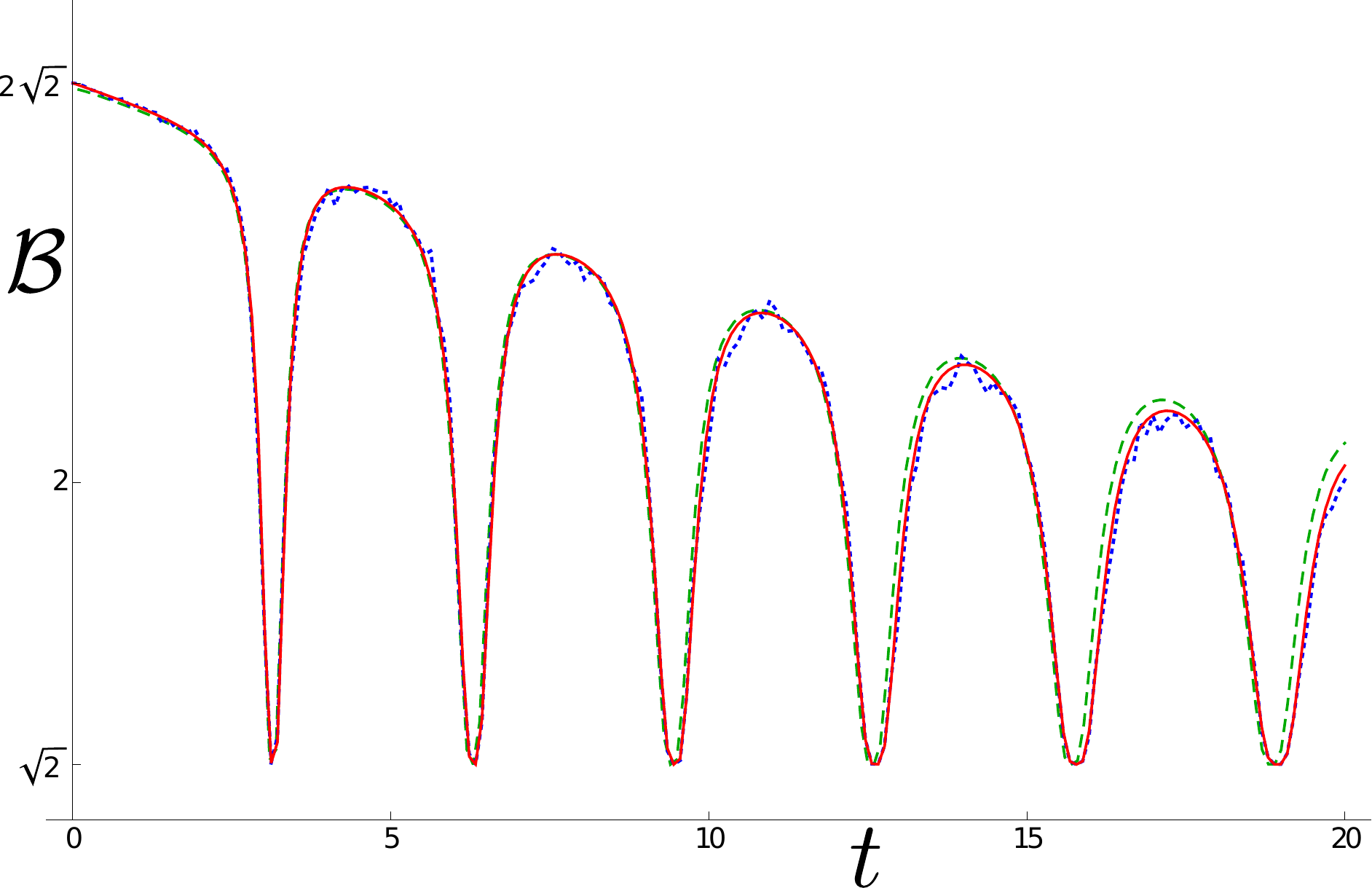}
\caption{$\mathcal B(t)~$ plotted for $\kappa = \text{0.05}~$ calculated by solution of Eq. (\ref{eqn:qsdeCSYS}) with a Hilbert space path integral with cutoff $c=2$ (green dashed), by solution of Eq. (\ref{eqn:linearizedcondmaster}) (red solid) and by averaging over an ensemble of 2000 quantum trajectories (blue dotted).} 
\label{fig:20diag005}
\end{figure}

\begin{figure}[h!tp]
\centering
\includegraphics[width=0.85\columnwidth]{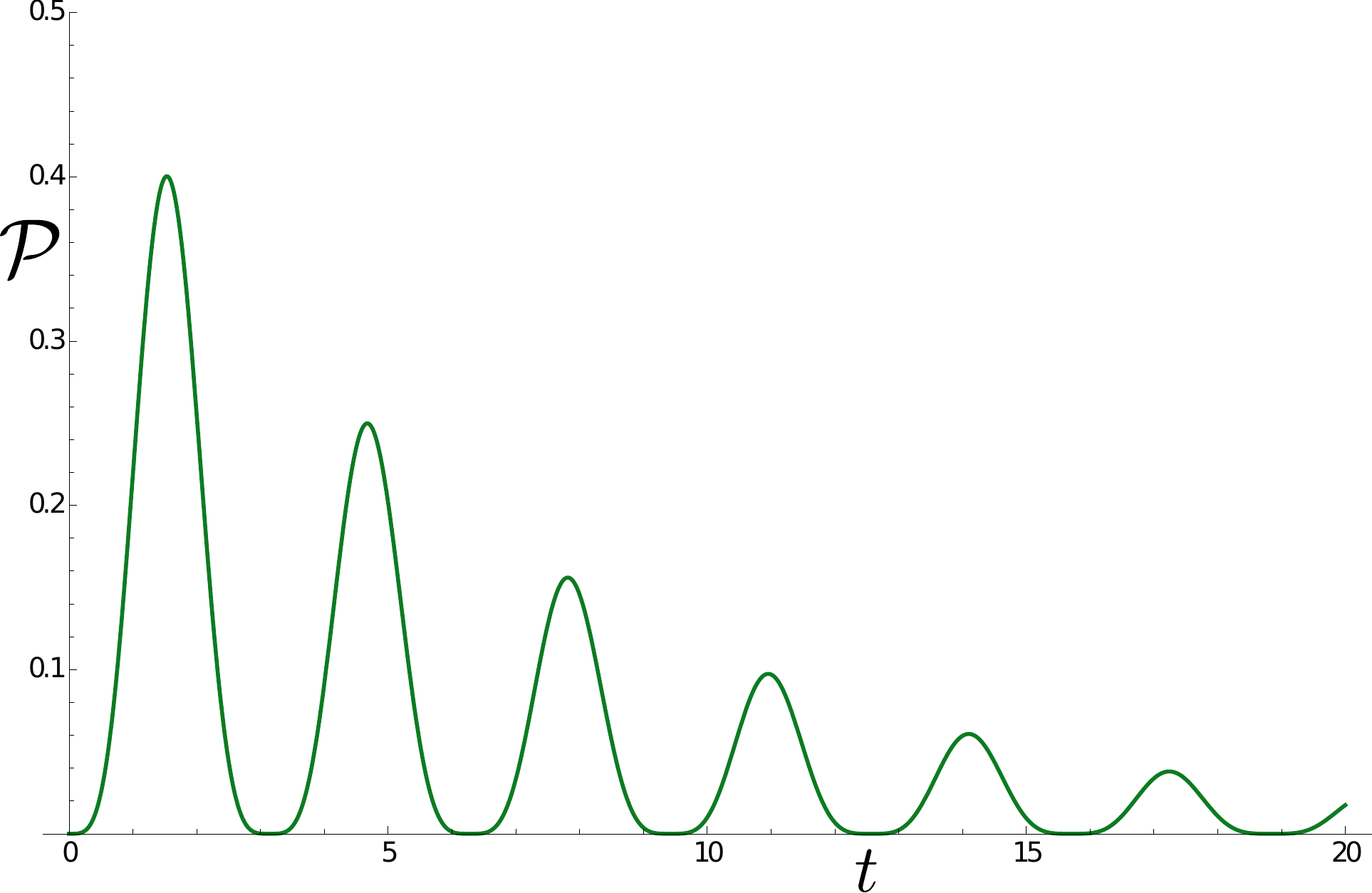}
\caption{Time dependence of the two photon emission probability $\mathcal P(t)~$ for $\kappa = \text{0.05}$. The distribution is not normalized, since only simultaneously emitted photons are considered. The plot displays Rabi-oscillations generated by the Jaynes-Cummings type coherent dynamics.} 
\label{fig:20diagProb005}
\end{figure}

\begin{figure}[h!tp]
\centering
\includegraphics[width=0.65\columnwidth]{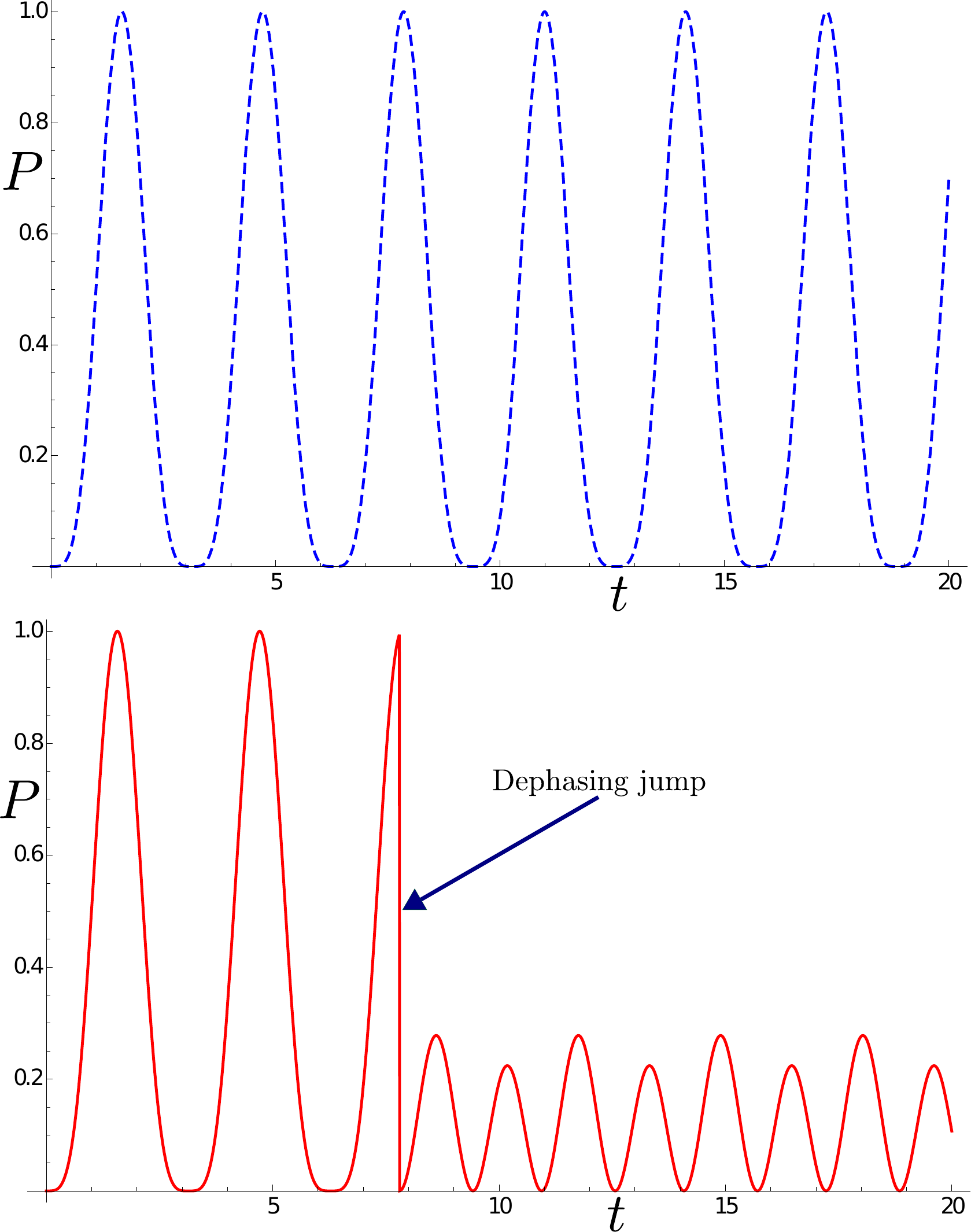}
\caption{Time dependence of the two photon population $P$~for a totally coherent trajectory (blue dashed) and a trajectory with one dephasing jump (red solid). The red curve shows clearly, that the phase coherence of the Rabi-oscillations in the blue plot gets lost due to dephasing.} 
\label{fig:trajectoriescompare}
\end{figure}

\begin{figure}[h!tp]
\centering
\includegraphics[width=0.85\columnwidth]{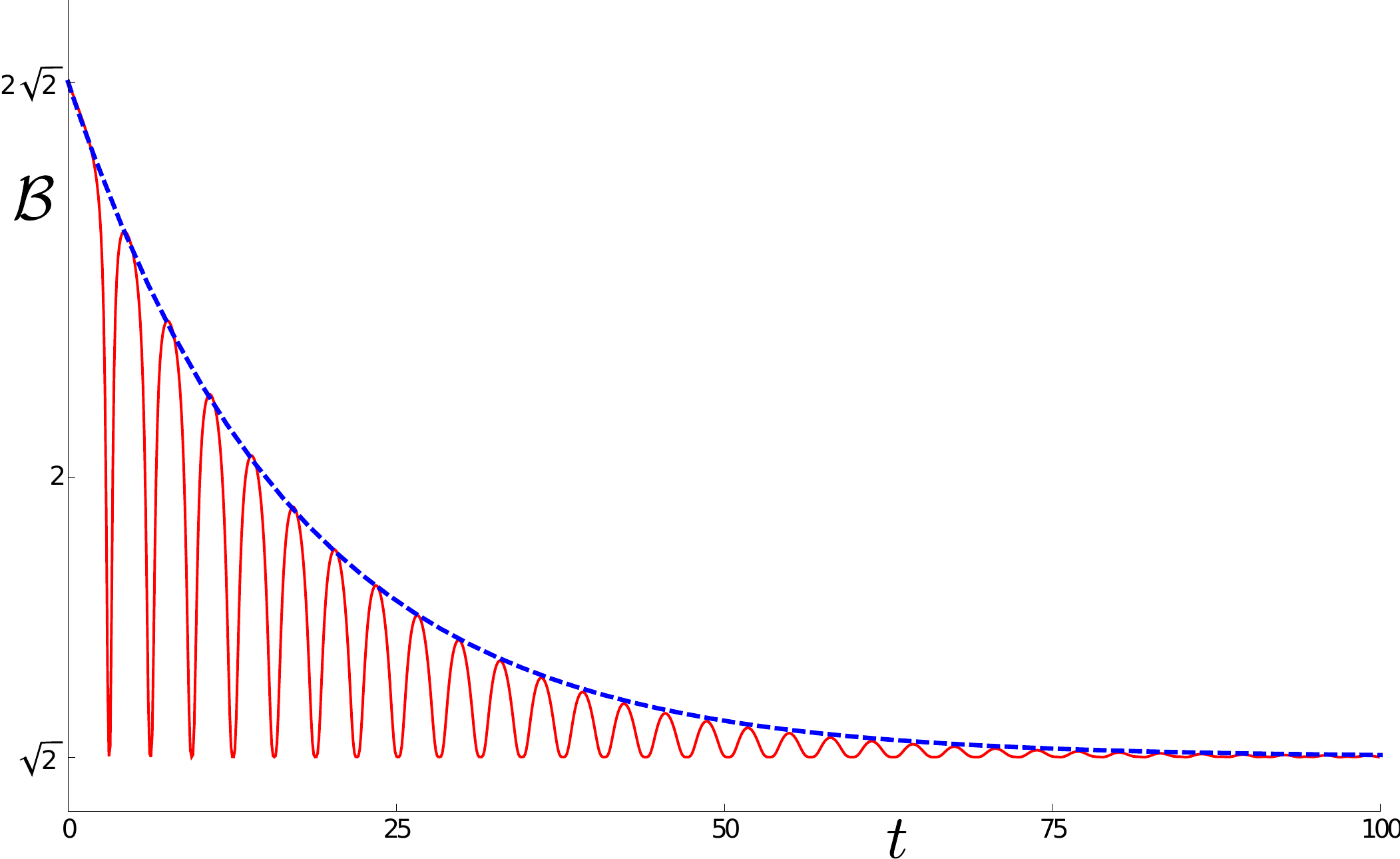}
\caption{Long-time decay of $\mathcal B(t)~$ plotted for $\kappa = \text{0.075 (red solid)}~$ and $f(t)= \sqrt{2}(1+\exp(-\text{0.585} t))~$ (blue dashed). The blue exponentially decreasing curve envelops the net decay of entanglement.} 
\label{fig:envelopefit}
\end{figure}

\begin{figure}[h!tp]
\centering
\includegraphics[width=0.85\columnwidth]{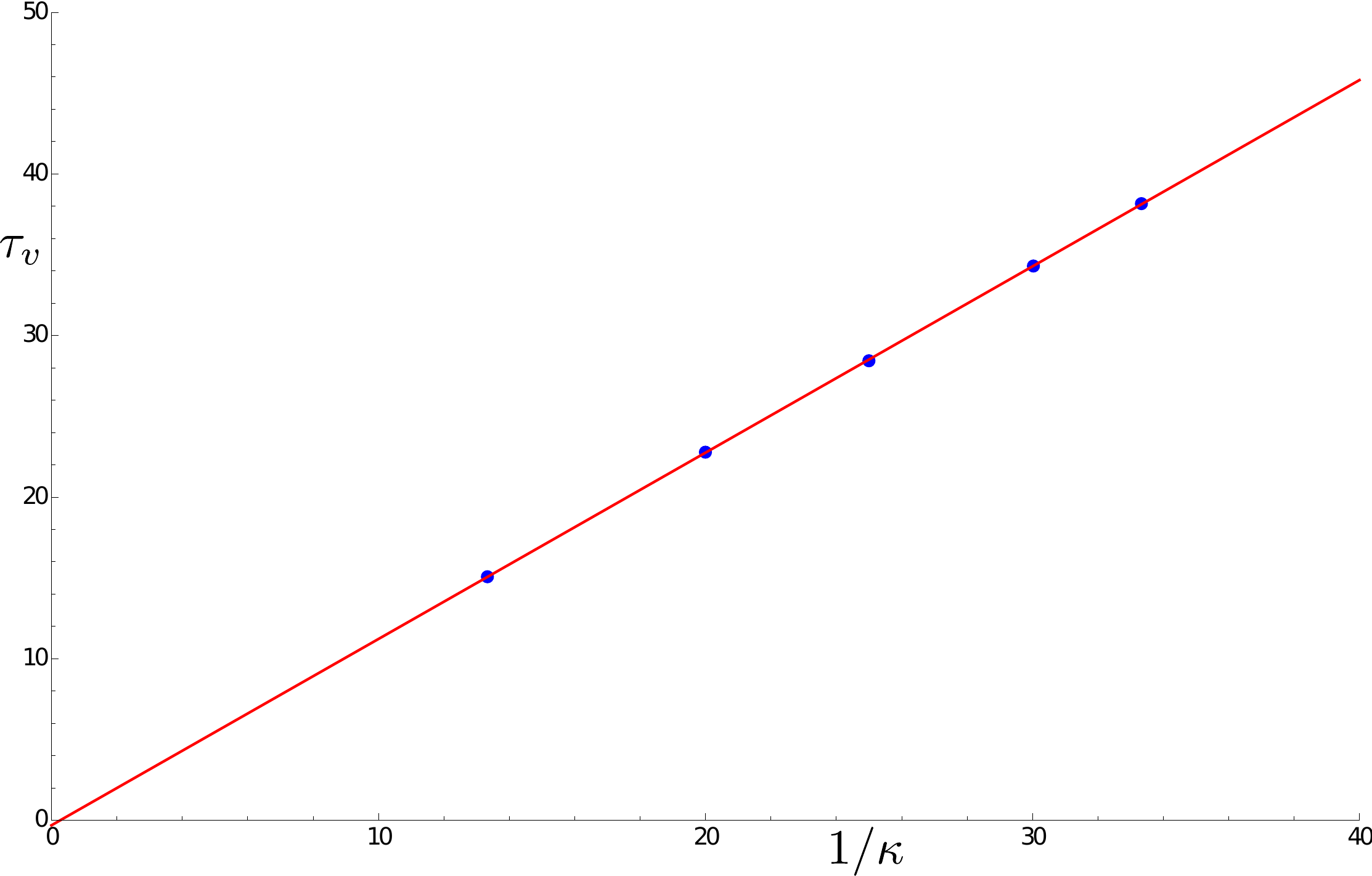}
\caption{The time $\tau_v~$ at which the CHSH violation vanishes plotted against $\frac{1}{\kappa}$ (blue points). The concurrence with the linear fit $t(1/\kappa) = \text{-0.3166} + \text{1.1529}\frac{1}{\kappa}~$ (red line) is on the spot.} 
\label{fig:belldependence}
\end{figure}

 The dephasing strength $\kappa$ is varied ranging from $\kappa = \text{0}~$ to $\kappa = \text{0.2}$. Without dephasing ($\kappa = \text{0}$) the expected value of $\mathcal B = 2 \sqrt{2}~$ can be reproduced for any given emission time. For a totally dephased product state, $\mathcal B = \sqrt{2}~$ holds for our measurement scheme presented in Section \ref{sec:chshtheory}. Hence, we expect in the presence of dephasing an asymptotic behaviour like $\mathcal B(t)\overset{t\rightarrow \infty}{\longrightarrow} \sqrt{2}$.\\
The time dependence $\mathcal B(t)~$ of the Bell parameter for $\kappa = \text{0.05}$~calculated with the three different integration techniques described in Section \ref{sec:chshtheory} is shown in Fig. \ref{fig:20diag005}.
The concurrence of the three plots is quite well, showing no systematic deviations between the numerically exact solution (red solid) and the noisy stochastic ensemble average (blue dotted). The path integral calculation with cutoff $c=2$~(green dashed) is almost congruent with the numerically exact solution for small emission times and shows significant aberrations systematically overestimating the entanglement only for $t>10$. That is because the weight of trajectories which are of higher order in $\kappa$~becomes significant for greater emission times so that we would need a higher cutoff to obtain adequate accuracy. All plots pictured in Fig. \ref{fig:20diag005} have two remarkable features. First, a net decay of $\mathcal B~$ due to dephasing, enveloping the oscillations can be found. Second, oscillations with the same frequency as those of the coherent emission probability (see Fig. \ref{fig:20diagProb005}) which is proportional to the population of the two photon states, i.e. $\mathcal P(t) \sim \langle \Psi(t)\rvert\left(\lvert 7\rangle \langle 7\rvert +\lvert 8\rangle\langle 8\rvert\right)\lvert \Psi(t)\rangle~$, are visible. 
The understanding of this complex time dependence lies at the heart of our quantum trajectory description and is one of the most interesting results of this work. The Jaynes-Cummings type dynamics brings about Rabi-oscillations between exciton and photon states in each of the optical dots which penetrate to the photon emission probability $\mathcal P~$ (see Fig. \ref{fig:20diagProb005}).
At the minima of $\mathcal P~$ the coherent dynamics forbids the emission of a photon because only the electron-HH states are populated and no photons are present. A dephasing event before the supposed measurement randomizes the phase of the coherent Rabi-oscillations. This enables trajectories which have been exposed to dephasing to emit photons at any instance in time, in particular at times where the coherent Rabi-oscillations enforced a purely excitonic state (compare the two plots in Fig. \ref{fig:trajectoriescompare}).
For these measuring times $t_m~$ the dephased trajectories contribute to a very large amount, up to hundred percent at the sharp minima of $\mathcal P$, which explains the minima of $\mathcal B(t)~$ with $\mathcal B(t_m)\approx \sqrt{2}~$ at these points.\\
The net decay of $\mathcal B(t)~$ is quite intuitive since the probability that a dephasing jump occurs increases monotonously with time.
To investigate the long-time behaviour of the entanglement decay it is obviously (see Fig. \ref{fig:envelopefit}) accurate to fit an exponential function to the net decay of entanglement.
The dependence of the exponential decay's measure on the dephasing strength is studied by plotting the time $\tau_v~$ at which $\mathcal B(\tau_v) = \text{2}~$ holds, that is to say at which the CHSH inequality is met, versus the inverse dephasing strength (see Fig. \ref{fig:belldependence}).
$\tau_v~$ increases linearly with $1/\kappa~$ which indicates that the measure of the exponential net decay is reciprocally proportional to the dephasing strength $\kappa$. For $1/\kappa \rightarrow \text{0},~\tau_v~$ goes up to a small error to zero, which displays infinitely fast entanglement decay for $\kappa \rightarrow \infty~$ as one would intuitively expect.
\subsubsection{Results for the interference approach}
\begin{figure}[h!tp]
\centering
\includegraphics[width=0.95\columnwidth]{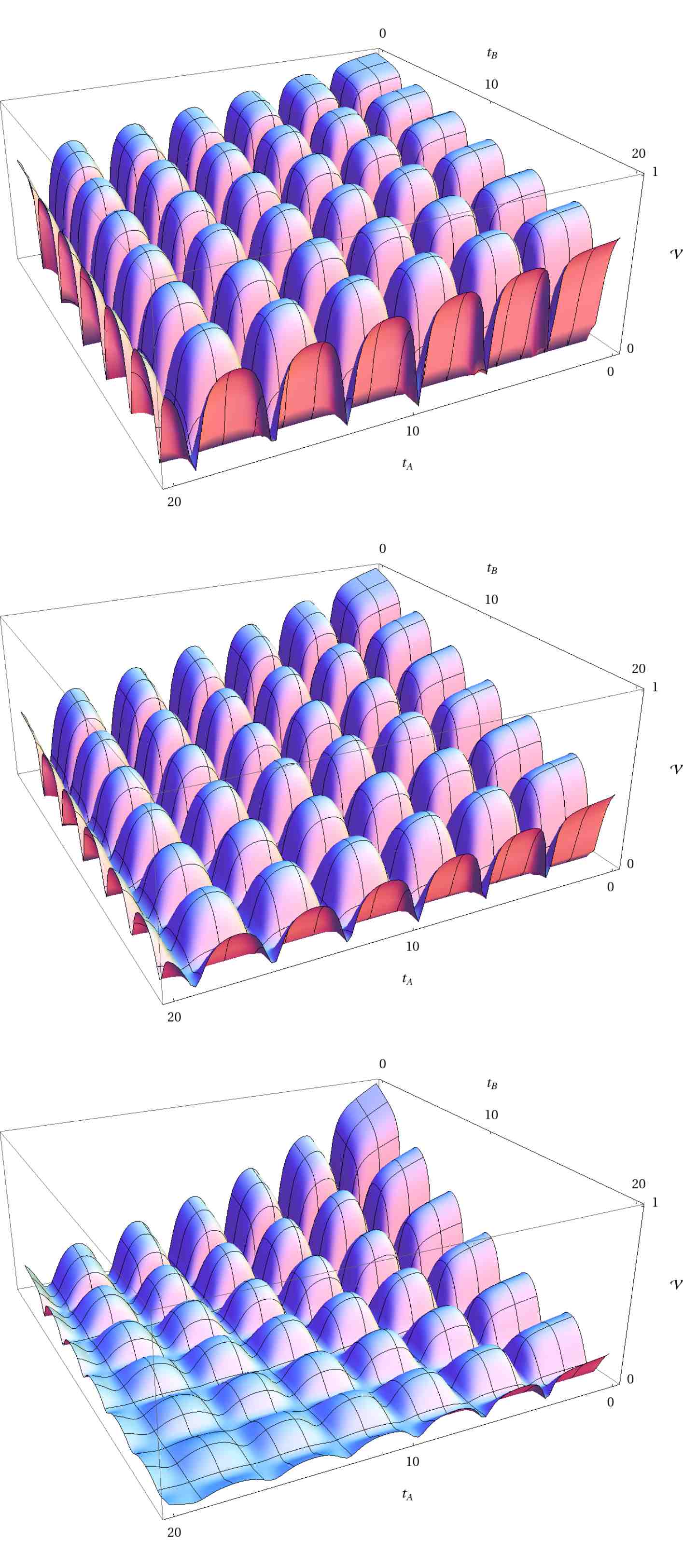}
\caption{Decay of $\mathcal V(t_B,t_A)~$ for $\kappa = \text{0.05 (top)},~\kappa = \text{0.10 (middle) and }\kappa = \text{0.20 (bottom)}$, calculated by solution of Eq. (\ref{eqn:linearizedcondmaster}). $\gamma = \text{0.1}~$ and $v_i=q_i= \text{1.0}~$ in all plots.} 
\label{fig:viscomparend}
\end{figure}

\begin{figure}[h!tp]
\centering
\includegraphics[width=0.85\columnwidth]{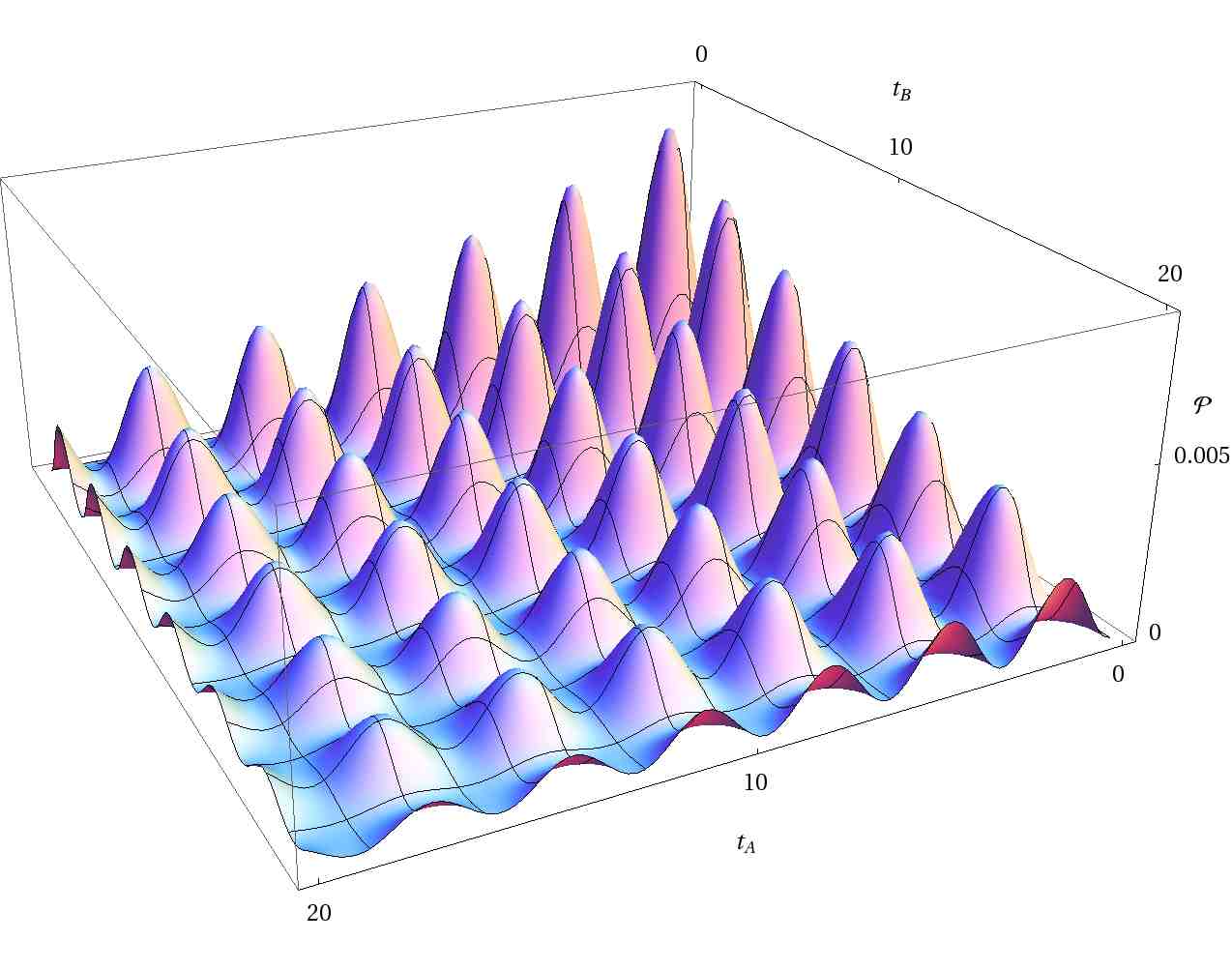}
\caption{$\mathcal P(t_B, t_A)~$ for $\gamma = \text{0.1 (top)}$, $\kappa = \text{0.05}~$ and $v_i=q_i= \text{1.0}$, calculated by solution of Eq. (\ref{eqn:linearizedcondmaster}).} 
\label{fig:probcomparend}
\end{figure}

\begin{figure}[h!tp]
\centering
\includegraphics[width=0.85\columnwidth]{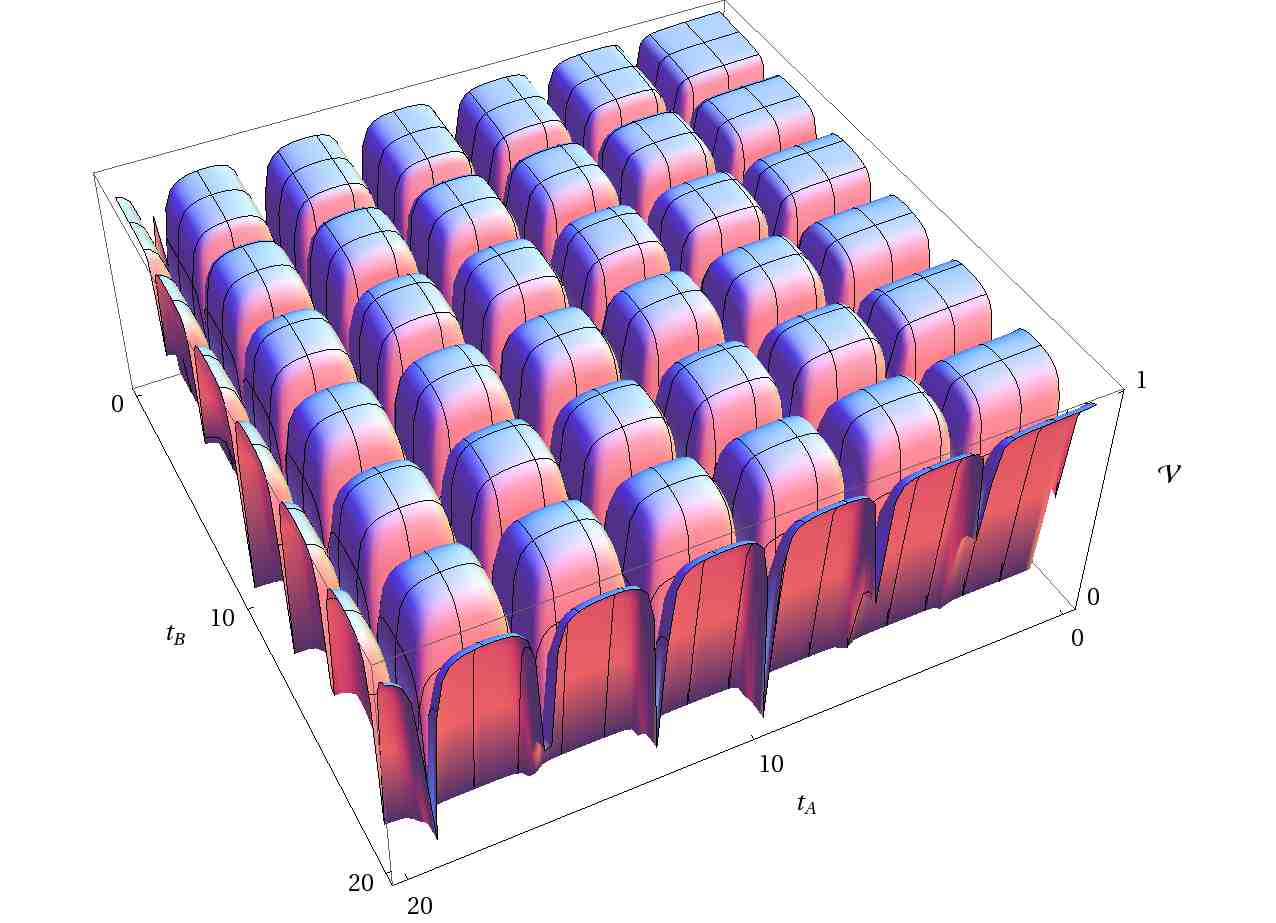}
\caption{Decay of $\mathcal V(t_B,t_A)~$ for $\kappa = \text{0.01}$, calculated by solution of Eq. (\ref{eqn:linearizedcondmaster}). $\gamma = \text{0.1}~$ and $v_i=q_i= \text{1.0}~$. The net decay is very slow but distinctive valleys are still visible for this weak dephasing strength.} 
\label{fig:weakdephasing}
\end{figure}

The decay of entanglement as measured by the visibility of interference fringes $\mathcal V(t_B,t_A)~$ is shown in Figure \ref{fig:viscomparend} for different dephasing strengths $\kappa = \text{0.05, 0.10, 0.20}$,  $\gamma = \text{0.1}~$ and $v_i=q_i= \text{1.0}$, calculated by numerical integration of Eq. (\ref{eqn:linearizedcondmaster}).
The behaviour is a straightforward generalization of the phenomenology discussed in Section \ref{sec:chshresults} to nonsimultaneous photon measurements. We again see a net decay with increasing emission times and characteristic oscillations following the Rabi-oscillations of the emission probability (see Fig. \ref{fig:probcomparend}).
The emission probability $\mathcal P~$ is defined in a straightforward way by 
\begin{align}
&\mathcal P(t_B,t_A) = \nonumber\\
&=\gamma^2\sum_{i,j=\pm}\text{Tr}\left\{\hat n_{\sigma_i^B}\tilde{\mathcal T}_{t_B,t_A}\left\{\hat n_{\sigma_j^A}\rho(t_A)\hat n_{\sigma_j^A}\right\}\hat n_{\sigma_i^B}\right\}\text{Tr}\{\tilde \rho(t_A)\}=\nonumber\\
& = \gamma^2(\tilde X+ \tilde Y),\qquad \text{w.l.o.g. }~t_A\le t_B,
\end{align}
where $\tilde X, \tilde Y~$ are the quantities defined in Eq. (\ref{eqn:xyzdef}) but with the unnormalized density matrix $\tilde \rho~$ introduced in Eq. (\ref{eqn:linearizedcondmaster}) used to calculate expectation values. $\text{Tr}\{\tilde \rho(t_A)\}~$ is the probability that no photon emission occurs before $t_A$.  $\tilde{\mathcal T}_{t_B,t_A}~$ denotes the time evolution operator governed by the flow of Eq. (\ref{eqn:linearizedcondmaster}) and brings about the conditional time evolution of the nonnormalized density matrix $\tilde \rho$ between $t_A$~and $t_B$.
As analyzed in the previous section we assume the oscillations of the visibility to disturb the functionality of the photon entangler only minimally, since the emission probability in the valleys of $\mathcal V(t_A,t_B)~$ is minimal. The major contribution to the mean value of the visibility,
\begin{equation}
\overline{\mathcal V} = \int_0^\infty d t_A\int_0^\infty d t_B \mathcal P(t_B,t_A)\mathcal V(t_B,t_A),
\end{equation}
is thus expected to be given by the net decay of $\mathcal V$. To confirm this hypothesis, we calculate and plot the visibility for a small dephasing strength $\kappa = \text{0.01}~$ for which the net decay of entanglement is very slow (see Fig. \ref{fig:weakdephasing}).
However, the oscillations bringing about distinctive valleys in the visibility suggest that the mean entanglement of the emitted photons could be significantly reduced. The mean visibility for this parameter set yields $\overline{\mathcal V} \approx \text{92.1}\%$ which justifies the statement that the oscillations do not disturb the functionality of the entangler to a large amount, because the loss of about eight percent is almost entirely covered by the weak net decay.\\
\subsection{Influence of asymmetries in the coherent couplings}
\label{sec:asymresults}
Now, we focus on the discussion of a different kind of non-ideality, namely the influence of asymmetries in the Jaynes-Cummings couplings $\{q_i,v_i\vert i=A,B\}~$ on the visibility $\mathcal V~$. From now on we neglect dephasing, i.e. we set $\kappa = \text{0}$.
\subsubsection{Asymmteries in the two-dot system}
\begin{figure}[h!tp]
\centering
\includegraphics[width=0.85\columnwidth]{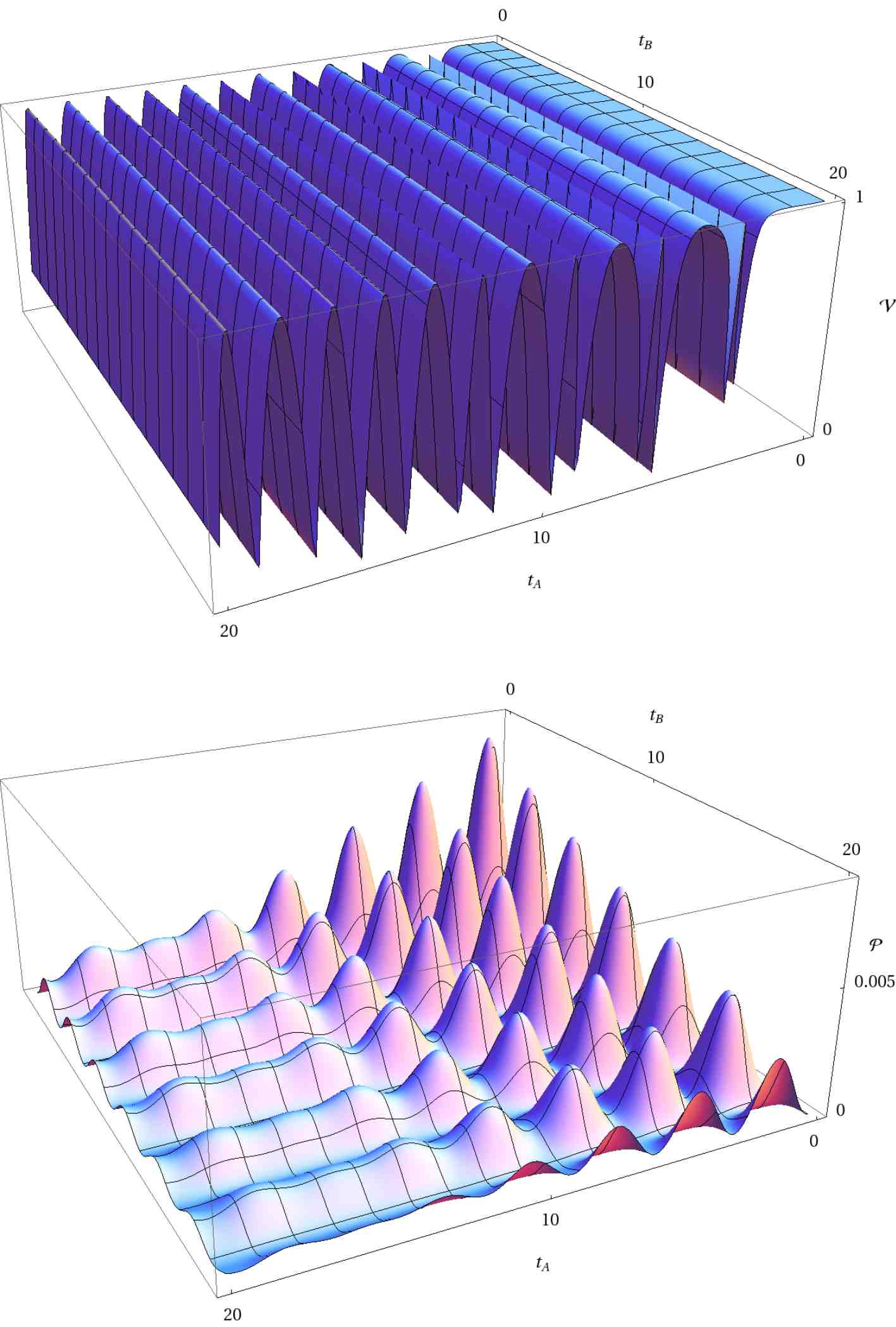}
\caption{$\mathcal V(t_B,t_A)~$(top) and $\mathcal P(t_B,t_A)~$ (bottom) for $v_A = \text{1.1},~ q_A = \text{1.0},~ v_B=q_B=\text{1.0}~$, calculated by solution of Eq. (\ref{eqn:linearizedcondmaster}). $\gamma = \text{0.1}~$ and $\kappa = \text{0}~$ in both plots.} 
\label{fig:plotssingleasym}
\end{figure}

We distinguish two types of imbalances, the first of which is an imbalance between A and B whereas the second asymmetry is due to unequal couplings for the different polarizations. A pure A-B-asymmetry, that is to say $v_A = q_A \ne v_B = q_B$, does not affect the entanglement of the emitted photons at all. That is because it does neither break the symmetry between the two kets $\lvert \sigma_A^-\sigma_B^+\rangle, \lvert\sigma_A^+\sigma_B^-\rangle$, the coherence between which makes up the entanglement, nor introduce any asymmetry between the two polarizations on single photon space. If we set up a polarization asymmetry by setting $v_A = \text{1.1},~ q_A = \text{1.0},~ v_B=q_B= \text{1.0}~$ only on subsystem A, this symmetry is broken causing a complex $t_A$-dependence of $\mathcal V$ whereas the visibility is independent of $t_B~$ (see Figure \ref{fig:plotssingleasym}).
The emission probability $\mathcal P~$ shows the regular Rabi-oscillations dependent on $t_B~$ whereas its $t_A$-dependence is more irregular, governed by the competing oscillation frequencies $v_A, q_A~$ (see Figure \ref{fig:plotssingleasym}).\\
\begin{figure}[h!tp]
\centering
\includegraphics[width=0.85\columnwidth]{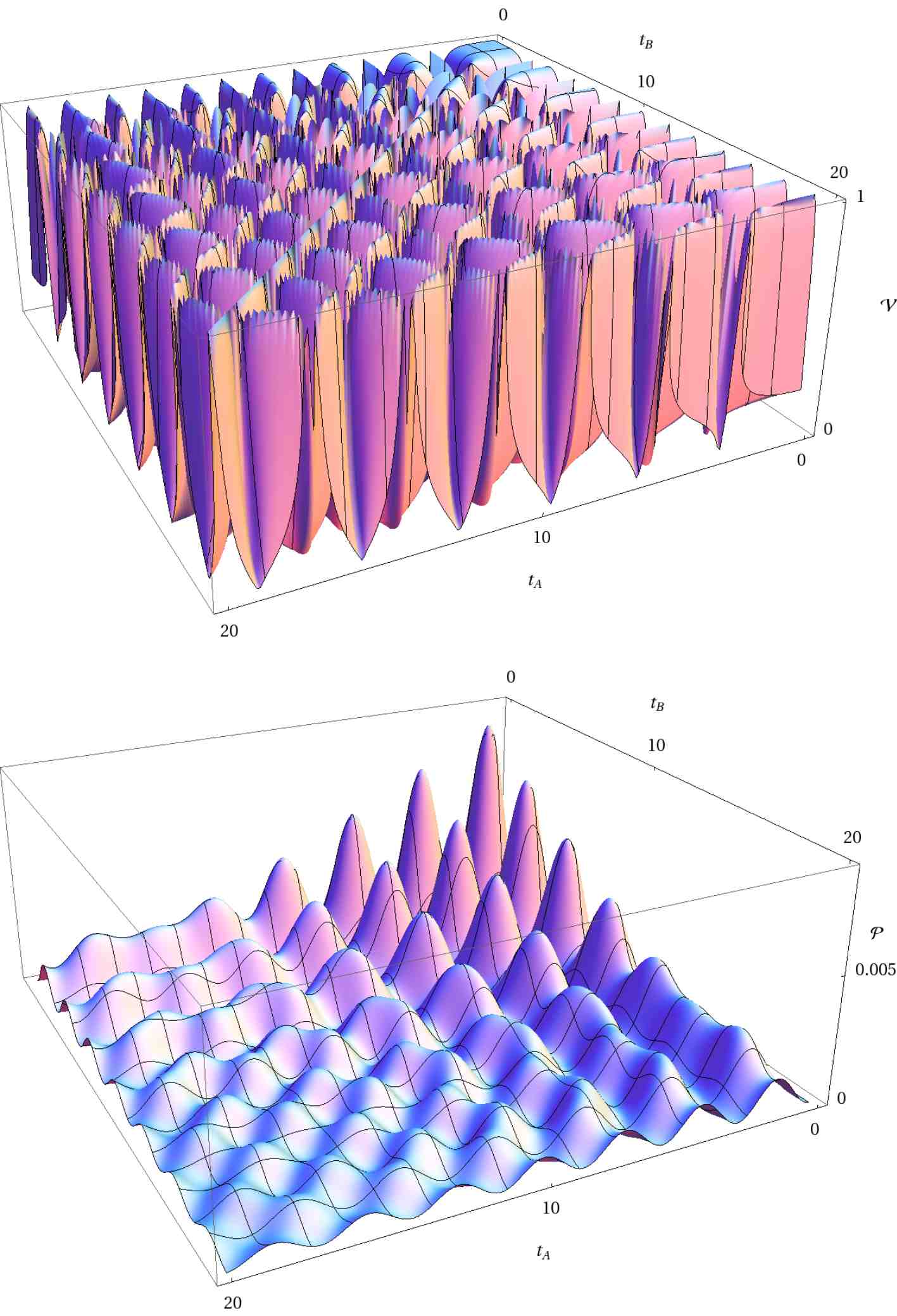}
\caption{$\mathcal V(t_B,t_A)~$(top) and $\mathcal P(t_B,t_A)~$ (bottom) for $v_A = v_B=\text{1.1},~ q_A =q_B=\text{1.0}$, calculated by solution of Eq. (\ref{eqn:linearizedcondmaster}). $\gamma = \text{0.1}~$ and $\kappa = \text{0}$ in both plots.} 
\label{fig:plotsdoubleasym}
\end{figure}
If a polarization asymmetry is introduced in the same way on both subsystems, for example by setting $v_A=v_B=\text{1.1},q_A=q_B=\text{1.0}$, the symmetry in the two excitation subspace of our Hilbert space (i.e. the symmetry between left and right hand side of the first row in Fig. \ref{fig:CSYSDynamics}) is not broken. However, the obvious asymmetry in the one exciton subspace will cause an effect as soon as one photon is emitted. If the two photons are emitted simultaneously $\mathcal V = \text{1}~$ thus holds (see the diagonal of the upper plot in Fig. \ref{fig:plotsdoubleasym}), while the behaviour of the visibility becomes rather irregular for arbitrary emission times $(t_A,t_B)$, nevertheless preserving $t_A$-$t_B$-symmetry (see the upper plot in Figure \ref{fig:plotsdoubleasym}). The emission probability (see the lower plot in Fig. \ref{fig:plotsdoubleasym}) exhibits oscillations with competing frequencies $v_i=\text{1.1},q_i=\text{1.0}~$ in both directions $t_A, t_B~$ also keeping the A-B-symmetry. To sum up the latter analysis as an advice to the experimentalists, we state that the $\sigma_+$-$\sigma_-$-symmetry of the cavity couplings should be the preliminary goal as to maximize the visibility. Furthermore, if the requirement of polarization-independence can not be met but A-B-symmetry is preserved, then the post selection rule $t_A=t_B~$ can guarantee perfectly entangled photons within the framework of our model. 
\subsubsection{Asymmetries in the tunnel couplings}
\begin{figure}[h!tp]
\centering
\includegraphics[width=0.85\columnwidth]{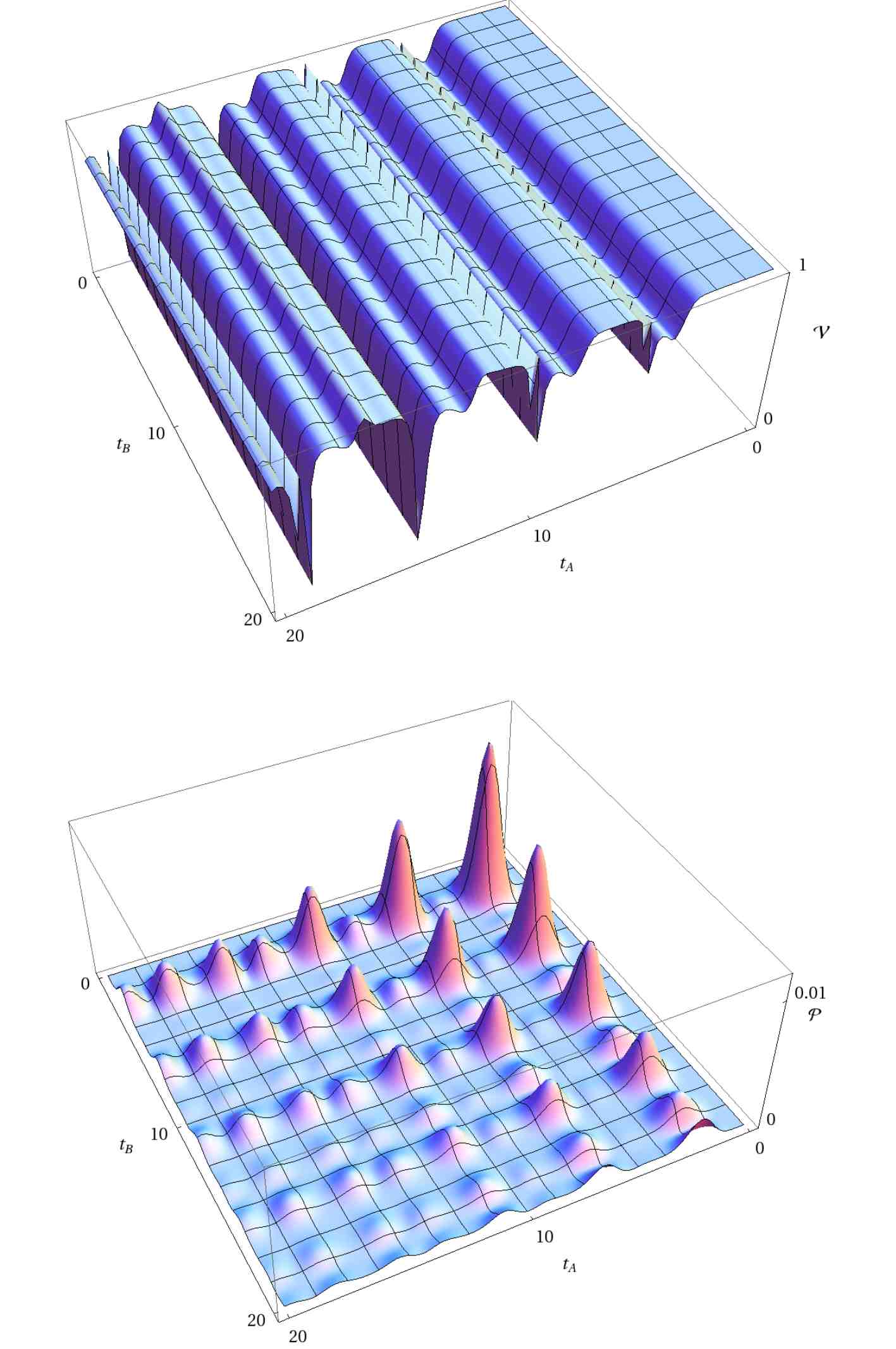}
\caption{$\mathcal V(t_B,t_A)~$(top) and $\mathcal P(t_B,t_A)~$ (bottom) for  $s_u^A= \text{1.1}\ne s_d^A = t_u^A = t_d^A = s_i^B=t_i^B= \text{1.0},~i=u,d$, calculated by solution of Eq. (\ref{eqn:linearizedcondmaster}). $\gamma = \text{0.2}~$ and $\kappa = 0$ in both plots. The cavity couplings $v_i=q_i= \text{1.0},~i=A,B~$ are chosen symmetrically.} 
\label{fig:coupledsinglepol}
\end{figure}

\begin{figure}[h!tp]
\centering
\includegraphics[width=0.85\columnwidth]{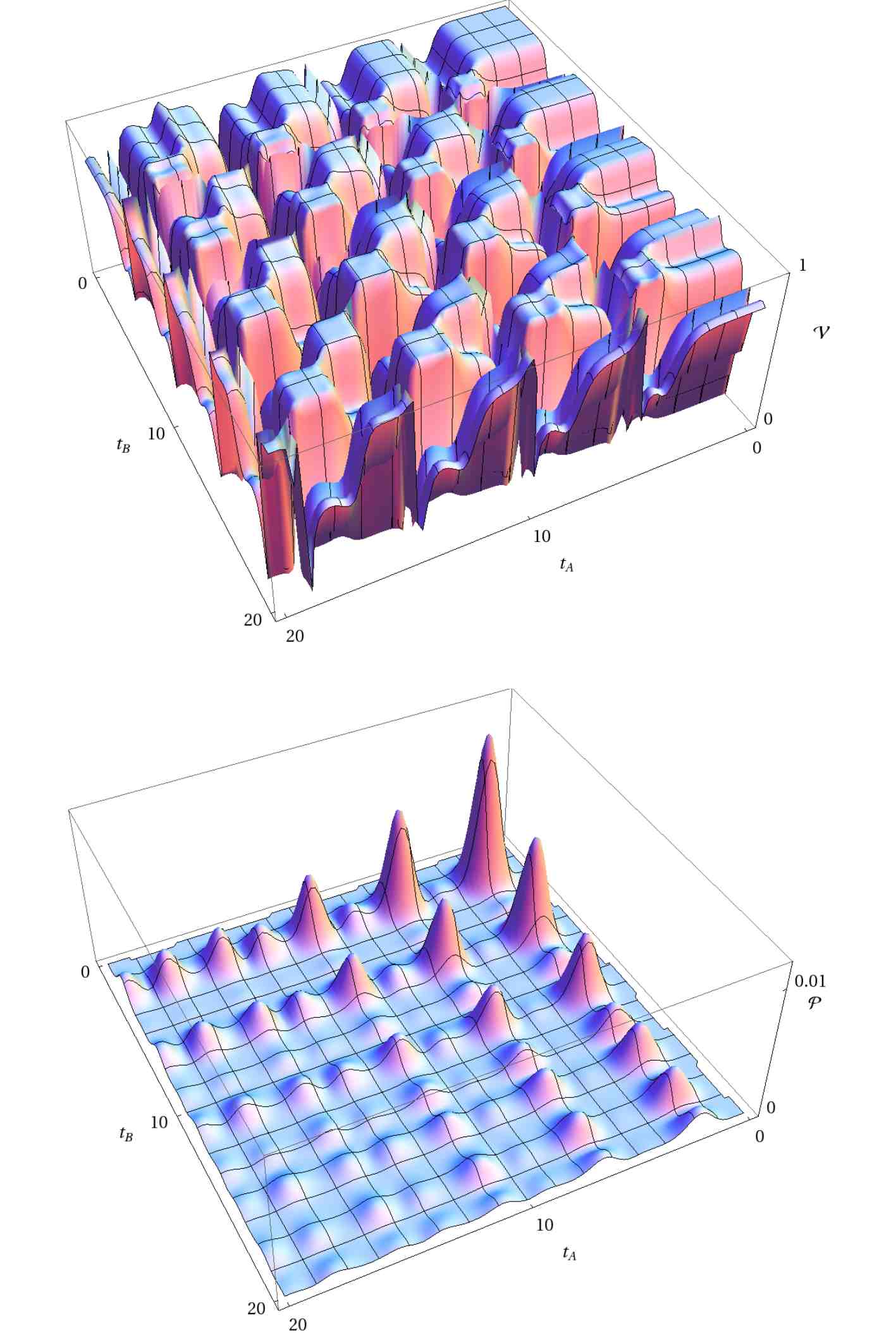}
\caption{$\mathcal V(t_B,t_A)~$(top) and $\mathcal P(t_B,t_A)~$ (bottom) for  $s_u^k= \text{1.1}\ne s_d^k = t_u^k = t_d^k= \text{1.0},~k=A,B$, calculated by solution of Eq. (\ref{eqn:linearizedcondmaster}). $\gamma = \text{0.2}~$ and $\kappa = \text{0}$ in both plots. The cavity couplings $v_i=q_i= \text{1.0},~i=A,B~$ are chosen symmetrically.} 
\label{fig:coupleddoublepol}
\end{figure}

In this part of the discussion we extend the coherent dynamics to the whole system's Hilbert space. That means we include the tunnel coupling to the lateral dots by considering the coupled Hamiltonian $H_S~$ (see Eq. (\ref{eqn:defhs})) as the generator of our coherent dynamics. Investigating the influence of asymmetries in the tunnel coupling strengths $\{s_i^j,t_i^j\vert i=u,d;\quad j= A,B\}~$ we observe a phenomenology similar to that occurring when consideration is given to imbalanced cavity couplings $\{v_i,q_i\vert i=A,B\}$ within the two-dot system. The idea of correcting the effect of these imbalances by systematically generating asymmetries in the tunnel couplings is therefore straightforward. Unfortunately, such a correction as measured by an improvement in mean visibility could not be achieved varying the relevant model parameters. Given symmetrical parameters for the two-dot system, the visibility again is inert regarding asymmetries which break A-B-symmetry but preserve electron-HH and polarization symmetry, that is $s_u^A=s_d^A=t_u^A=t_d^A\ne s_u^B=s_d^B=t_u^B=t_d^B~$. The visibility also stays at its maximal value $\mathcal V = \text{1}~$ if we break the electron-HH symmetry by setting $s_u^k=s_d^k \ne t_u^k = t_d^k,~k=A,B$. This observation displays the fact that we have to introduce an asymmetry which is polarization selective, if we want to touch the polarization entanglement of the emitted photons. The simplest way to do that is by choosing $s_u^k\ne s_d^k = t_u^k = t_d^k~$ on one dot k. The resulting visibility $\mathcal V(t_B,t_A)~$ and emission probability $\mathcal P(t_B,t_A)~$ are shown in Figure \ref{fig:coupledsinglepol}.
As expected, the visibility is only dependent on $t_A~$, since the polarization asymmetry concerns only subsystem A. The emission probability exhibits the same behaviour as in the symmetrical case concerning its $t_B$-dependence, whereas its $t_A$-dependence is influenced by the two competing coupling frequencies $s_u^A\ne s_d^A$.\\
Let us now create a polarization asymmetry which preserves A-B-symmetry by setting  $s_u^k\ne s_d^k = t_u^k = t_d^k~$ on both dots k = A,B. Like for the two-dot system we then expect the coherence of our relevant states to be untouched as long as both excitations are present in the system. The asymmetry becomes only relevant after the first photon has been measured. The visibility should therefore be maximal for $t_A=t_B$. This behaviour, together with the expected A-B-symmetry of $\mathcal V,\mathcal P~$ can be seen in Figure \ref{fig:coupleddoublepol}.\\
The concluding suggestion to the experimentalist striving for the measurement of entangled photons is very similar to the discussion of asymmetries in the Jaynes-Cummings couplings but now with the tunnel couplings $\{s_i^k,t_i^k\}~$ in the roll of $\{v_k,q_k\}$.
\section{Conclusions}
\label{sec:conclusions}
This final section is aimed to review the most important findings and to give an outlook as to future research which may be built on this work.\\ 
The first remarkable observation is the appearance of characteristic oscillations in the emission time dependent decay of two photon entanglement which are congruent with the Rabi-oscillations of the emission probability $\mathcal P$. The understanding of this phenomenon has required us to employ a quantum trajectory picture dividing the ensemble of quantum trajectories into two classes: The first class contains all realizations which have not been exposed to dephasing before the final photon emission whereas trajectories belonging to the second class have been dephased. The fingerprint of this unravelling in terms of quantum trajectories visible in the results for the ensemble average justifies the formally involved treatment of the open system's dynamics in terms of a PDP in the reduced system's Hilbert space. Furthermore, it has been shown that the long time decay of entanglement follows indeed an exponential decay which is a typical behaviour with Markovian dissipation. In this context the reciprocal proportionality of the measure of this decay to the dephasing strength $\kappa~$ has been demonstrated. Asymmetries in both the coherent couplings of the electron-HH excitations to the cavity mode and the tunnel couplings between optical and lateral dots have been investigated without dephasing. The most important result of these simulations is that polarization-symmetry of the respective quantities is the crucial point concerning reliable production of entangled photons. However, if this polarization-symmetry can not be assured, A-B-symmetry of the system's Hamiltonian can also guarantee perfect entanglement if only simultaneously emitted photons are considered.\\

In the future, it might be interesting as well as experimentally relevant to extend our analysis to the non-Markovian regime which would allow the modelling of further dissipation channels, e.g. hyperfine interaction. The necessity of such a treatment depends on the particular dephasing channel that is the most relevant one of a given host material for the lateral and optical quantum dots.\\

\section*{Acknowledgments}
We would like to thank Thomas Balder and Leo Kouwenhoven who contributed to early stages of this work and acknowledge financial support by the German Science Foundation (DFG).
\bibliography{entanglerpdflatex}
\bibliographystyle{prsty}
\end{document}